\documentclass{emulateapj}
\usepackage{color}
\usepackage{rotating}
\usepackage{gensymb}

\newcommand{\vecr}{\mbox{\boldmath $r$} {}}
\newcommand{\vecJ}{\mbox{\boldmath $J$} {}}
\usepackage{epsfig}
\usepackage{amsmath}
\newif\ifAMStwofonts
\shorttitle{Low-mass bodies in retrograde disks}
\shortauthors{S\'anchez-Salcedo et al.}
\begin{document}
\title{Torques on low-mass bodies in retrograde orbit in gaseous disks}
\author{F. J. S\'anchez-Salcedo\altaffilmark{1}, Ra\'ul O. Chametla\altaffilmark{2}, and A. Santill\'an\altaffilmark{3}}
\altaffiltext{1}{Instituto de Astronom\'{\i}a, Universidad Nacional Aut\'onoma de M\'exico, A. P. 70-264,
Mexico City 04510, Mexico \email{(jsanchez@astro.unam.mx)}}
\altaffiltext{2}{Universidad Aut\'onoma de Nuevo Le\'on, Facultad de Ciencias F\'{\i}sico-Matem\'aticas,
San Nicol\'as de la Garza, 66451, N. L., Mexico}
\altaffiltext{3}{Direcci\'on General de C\'omputo y Tecnolog\'{\i}as de la Informaci\'on y la
Comunicaci\'on (DGTIC), Universidad Nacional Aut\'onoma de M\'exico, Mexico City, Mexico}


\begin{abstract}
We evaluate the torque acting on a gravitational perturber 
on a retrograde circular orbit in the midplane of a gaseous disk.
We assume that the mass of this satellite is so low it weakly disturbs
the disk (type I migration). The perturber may represent the companion
of a binary system with a small mass ratio. 
We compare the results of hydrodynamical simulations with analytic predictions.
Our two-dimensional (2D) simulations indicate that the torque acting on
a perturber with softening radius $R_{\rm soft}$ can be accounted
for by a scattering approach if $R_{\rm soft}<0.3H$, where 
$H$ is defined as the ratio between the sound speed and the angular
velocity at the orbital radius of the perturber. For $R_{\rm soft}>0.3H$,
the torque may present large and persistent oscillations, but
the resultant time-averaged torque decreases rapidly with increasing $R_{\rm soft}/H$,
in agreement with previous analytical studies.
We then focus on the torque acting on small-size perturbers embedded
in full three-dimensional (3D) disks and argue that the
density waves propagating at distances $\lesssim H$ from the perturber contribute
significantly to the torque because they transport angular momentum.
We find a good agreement between the torque found in 3D simulations and analytical
estimates based on ballistic orbits. We compare the radial migration timescales of prograde
versus retrograde perturbers.
For a certain range of the perturber's mass and aspect ratio of the disk, 
the radial migration timescale in the retrograde case 
may be appreciably shorter than in the prograde case. We also provide the
smoothing length required in 2D simulations in order to account for 3D effects.

\end{abstract}


\section{Introduction}
\label{sec:intro}

Black hole (BH) binaries in the centers of galaxies may lose orbital angular momentum
as a result of the gravitational interaction with the surrounding gas and stars 
\citep[e.g.,][]{iva99,mil01,arm02,esc04,cua09,tan09,koc12,may13},
which leads to a reduction of the 
separation between the BHs. In merging galaxies, the 
strong gas inflows may induce the formation of a circumbinary rotating disk. 
Tidal torques between the binary and the
circumbinary disk may play an important role in the orbital evolution
of the binary.


In the center of galaxies,  BH binaries do not necessarily corotate with the disk because
the angular momentum of new material accreted into the circumbinary disk may have a different
(random) direction. \citet{nix11} suggest that retrograde flows are as likely as prograde
\citep[see also][]{nix12,mil13}.
Interestingly, \citet{nix11}  find that  for retrograde binaries with mass ratios $\geq 0.1$, 
mass capturing into the secondary BH near apocenter leads to
a rapid decrease in the orbital angular momentum of the binary while only a modest change in its
orbital energy; the eccentricity of the binary grows and the pericenter distance is shortened. 
This reduction of the pericenter distance may bring the BHs close enough for
gravitational wave radiation to drive their merger.

A number of papers have studied the evolution of the orbital parameters
of a binary surrounded by a retrograde accretion disk using numerical
simulations \citep[e.g.,][]{roe14,ban15,nix15,ama16}.
All the abovementioned papers consider 
the non-linear regime, that is, they adopt secondary to primary mass ratios $q$ 
between $0.1$ and $1$.

On the other hand, \citet{iva15} study a binary with small $q$ immersed in a retrograde
disk through semi-analytic methods and two-dimensional (2D) simulations.
They mainly focus on binary mass ratios that are small
but large enough to significantly perturb the disk.

Here we study the accretion and gravitational torques acting on a 
retrograde satellite with such a small mass that it is unable to carve a gap
or open a cavity in the disk. In the context of planetary migration,
this regime is referred to as type I migration. 
The perturbing mass may represent the 
secondary component of a binary with small mass ratio, or a stellar cluster
or dwarf galaxy embedded in a gas-rich host galaxy. In this paper, we will 
mainly talk about binary systems, but the majority of our results can be 
translated to galactic systems.   

Binary systems consisting of two BHs and having small values of $q$ can
arise in the core of the merger of unequal-mass galaxies \citep[e.g.,][]{kha12}.
They can also arise in systems 
composed by a supermassive BH and a stellar-mass BH (or star) trapped in 
the accretion disk \citep[e.g.,][]{bar17}, or in systems composed
by a supermassive BH and an intermediate-mass BH.
In fact, it has been suggested that intermediate-mass BHs can
grow efficiently in disks around supermassive BHs via mass accretion
\citep{mck12}.

The paper is organized as follows. A short overview of the assumptions together with some definitions 
are given in Section \ref{sec:assumptions}. 
In Section \ref{sec:excitation_torque}, we compile the different analytical estimates of the gravitational 
torque acting on a retrograde satellite embedded in a gaseous disk.
In Section {\ref{sec:hydro_sims}, these analytical estimates are compared to the results based on hydrodynamical
simulations. Some implications of our results are presented in Section \ref{sec:discussion}. 
Our conclusions are given in Section \ref{sec:conclusions}.

\section{Basic assumptions and relevant parameters}
\label{sec:assumptions}
Consider a rotating disk with surface density $\Sigma(\vecr,t)$ in a spherical background potential
$\Phi_{b}$. Suppose that a small satellite (perturber) of mass $M_{2}$ 
is on a fixed circular orbit with radius $a_{2}$. The mass within 
a sphere of radius $a_{2}$ is $M_{1}$. The ratio between these masses,
$q\equiv M_{2}/M_{1}$, is assumed to be small. Therefore,
the perturber rotates with angular velocity $\omega\equiv \sqrt{GM_{1}/a_{2}^{3}}$.
In the particular case of a binary, $M_{1}$ and $M_{2}$ are the masses of the primary and
the secondary, respectively.
Note that by assumption, the disk's mass is negligible compared to $M_{1}$.

We assume that the interaction between the perturber and the disk is exclusively gravitational;
the perturber does not have any rigid surface.
The gravitational potential associated with the perturber is given by
\begin{equation}
\Phi_{2}(\vecr,t)=-\frac{GM_{2}}{\sqrt{|\vecr-\vecr_{2}|^{2}+R_{\rm soft}^{2}}},
\end{equation}
where $\vecr_{2}(t)$ is the position of the perturber and $R_{\rm soft}$ 
is a softening radius, which is nonzero for finite-sized bodies and it is zero 
for point-mass accretors.  
The exchange of angular momentum with the disk through 
waves produces a dynamical torque on the perturber.
In addition to this dynamical torque, and for point-like satellites ($R_{\rm soft}=0$),
accretion of material may lead to an ``aerodynamic'' drag.

For systems with a mass ratio $q$ well below a certain critical value $q_{\rm cr}$, 
the torques expressed on the disk are too small to significantly alter its density structure
(type I migration). 
However, for $q\gtrsim q_{\rm cr}$, the torques can 
open an annular gap in the surface density of the
disk at the location of the perturber (type II migration).

In the prograde case, $q_{\rm cr}$ is given by
\begin{equation}
q_{\rm cr}^{(p)}= 10 \left(\frac{\nu}{\omega a_{2}^{2}}\right)^{1/2} h^{3/2}=10\sqrt{\alpha}h^{5/2},
\label{eq:qcr_prograde}
\end{equation}
where $\nu$ is the viscosity of the disk, $\alpha$ is the Shakura-Sunyaev viscosity parameter 
and $h$ is the disk's aspect ratio $h\equiv H/a_{2}$ \citep[e.g.,][]{lin86,cha17}.
Here $H$ is the vertical scaleheight of the disk at the perturber's location.
Along the paper, we will use the scripts $(p)$ and $(r)$ to 
distinguish between the prograde case and the retrograde case.

In the retrograde case and assuming that the perturbing particle is a perfect
accretor, such as a BH or a star, $q_{\rm cr}^{(r)}$ can be obtained by imposing that
its mass accretion rate must be smaller than the radial flow rate of
mass in the disk. For a Keplerian disk with viscosity $\nu$, 
the radial mass flow is $3\pi \nu\Sigma$, while the accretion rate by
the perturber is $\pi \rho R_{\rm acc}^{2}V_{\rm rel}$, where $\rho$ is the disk's density at 
the cylindrical radius $a_{2}$,
$V_{\rm rel}$ is the relative velocity of the
perturber with respect to the gas and $R_{\rm acc}\equiv 2GM_{2}/V_{\rm rel}^{2}$ is the accretion radius.
In the retrograde case, $V_{\rm rel}=2\omega a_{2}$ and $R_{\rm acc}=qa_{2}/2$.
For the density, we take $\rho=\Sigma/(\sqrt{2\pi} H)$, which
corresponds to the midplane density of a Gaussian vertical profile.
Hence, the type I condition implies
\begin{equation}
q< q_{\rm cr}^{(r)}\equiv 4 \left(\frac{\nu}{\omega a_{2}^{2}}\right)^{1/2}h^{1/2}.
\label{eq:qcr_ac}
\end{equation}
In terms of the Shakura-Sunyaev viscosity parameter $\alpha$,
$q_{\rm cr}^{(r)}$ can be expressed as $q_{\rm cr}^{(r)}= 4\alpha^{1/2} h^{3/2}$.
The type I condition can be recast in terms of the accretion radius as
\begin{equation}
R_{\rm acc}\lesssim 2\left(\frac{\nu}{\omega a_{2}^{2}}\right)^{1/2}
h^{1/2}a_{2}.
\end{equation}
For a retrograde disk with parameters $h=0.05$ and $\nu=10^{-5}\omega a_{2}^{2}$,
the type I condition is fulfilled as long as 
$q\lesssim q_{\rm cr}^{(r)}=2.8\times 10^{-3}$ or, equivalently, $R_{\rm acc}\lesssim 0.028H$.

\citet{iva15} evaluate the minimum value of $q$ for gap opening through tidal torques,
instead of being the result of depletion by accretion.
They establish that tidal torques are too weak to open a gap if $q\leq 1.57h^{2}$. This
upper limit for $q$ coincides with our condition (\ref{eq:qcr_ac})
for a disk with $h=0.05$ and $\nu=2\times 10^{-5} \omega a_{2}^{2}$. For lower viscosities or/and
thicker disks, the condition given in Eq. (\ref{eq:qcr_ac}) is more stringent.

Retrograde systems with $q$ larger than $q_{\rm cr}^{(r)}$ may still lie in the type I regime 
if the perturber is an extended body. In particular, the type I condition is met 
if $R_{\rm soft}\gg R_{\rm acc}$ because the response of the disk is linear 
at any location in this case \citep[e.g.,][]{ber13,iva15}.

\section{Torques in a retrograde disk}
\label{sec:excitation_torque}
Without loss of generality, we take in both theoretical analysis and simulations throughout 
the paper, that the disk rotates in counterclockwise direction. 
For the retrograde case, the satellite rotates in the clockwise direction. 
We denote $dT_{g}/dR$ the torque acting upon an elementary ring of radius $R$ and width 
$\delta R$, and use the sign convention that the ring loses (gains) angular momentum
if $dT_{g}/dR$ is negative (positive).
\subsection{Strict 2D disk}

A first estimate of the excitation torque density $dT_{g}/dR$ exerted by
a retrograde satellite on a 2D disk can be obtained in the impulse approximation
in a similar way as Lin \& Papaloizou (1979, 1993) did for the prograde case. 
The impulse approximation yields
\begin{equation}
\frac{dT_{g}}{dR}\simeq - \frac{1}{2}q^{2} \Sigma a_{2}^{3}\omega^{2} \left(\frac{a_{2}}{R-a_{2}}\right)^{2}.
\label{eq:impulse_app1}
\end{equation}
\citet{iva15} use that expression for the torque density to predict the density profile of 
the disk and the gap profile in the nonlinear case.

A more refined derivation of the excitation torque density can be done using 
perturbation analysis of the orbits (see Appendix). For a Keplerian disk,
we find
\begin{equation}
\frac{dT_{g}}{dR}=-\frac{1}{8}q^{2}\Sigma \omega^{2} a_{2}^{3} \left(\frac{a_{2}}{R}\right)^{3}
\left(-2K_{0}\left[\frac{\xi}{2R}\right]+
\frac{\Delta}{\xi}K_{1}\left[\frac{\xi}{2R}\right]\right)^{2}
\label{eq:torqueKK}
\end{equation}
where $\Delta\equiv R-a_{2}$, $\xi=(\Delta^{2}+R_{\rm soft}^{2})^{1/2}$ and $K_{n}$ are
the modified Bessel functions.
For impact parameters such that $\xi \ll R $, Equation (\ref{eq:torqueKK}) can be approximated by
\begin{equation}
\frac{dT_{g}}{dR}\simeq -\left(\frac{1}{2}\right) q^{2} \Sigma \omega^{2} a_{2}^{3}
\left[\frac{R \Delta}{\xi^{2}}+ \ln\left(\frac{\xi}{4R}\right)+0.577\right]^{2}.
\label{eq:Tg_orbital_scatter}
\end{equation}
In general, the impulse approximation (Equation \ref{eq:impulse_app1}) overestimates (underestimates)
the torque density in the external (internal) disk, as compared to Equation (\ref{eq:Tg_orbital_scatter}).
Only in the particular case that  $R_{\rm soft}\ll |\Delta|\leq 0.05a_{2}$, we can approximate the torque density 
by Equation (\ref{eq:impulse_app1})
with an error $\lesssim 50\%$ in the external disk ($\Delta >0$) and $\lesssim 30\%$ in the internal
disk ($\Delta <0$).

\begin{figure}
\includegraphics[width=88mm,height=65mm]{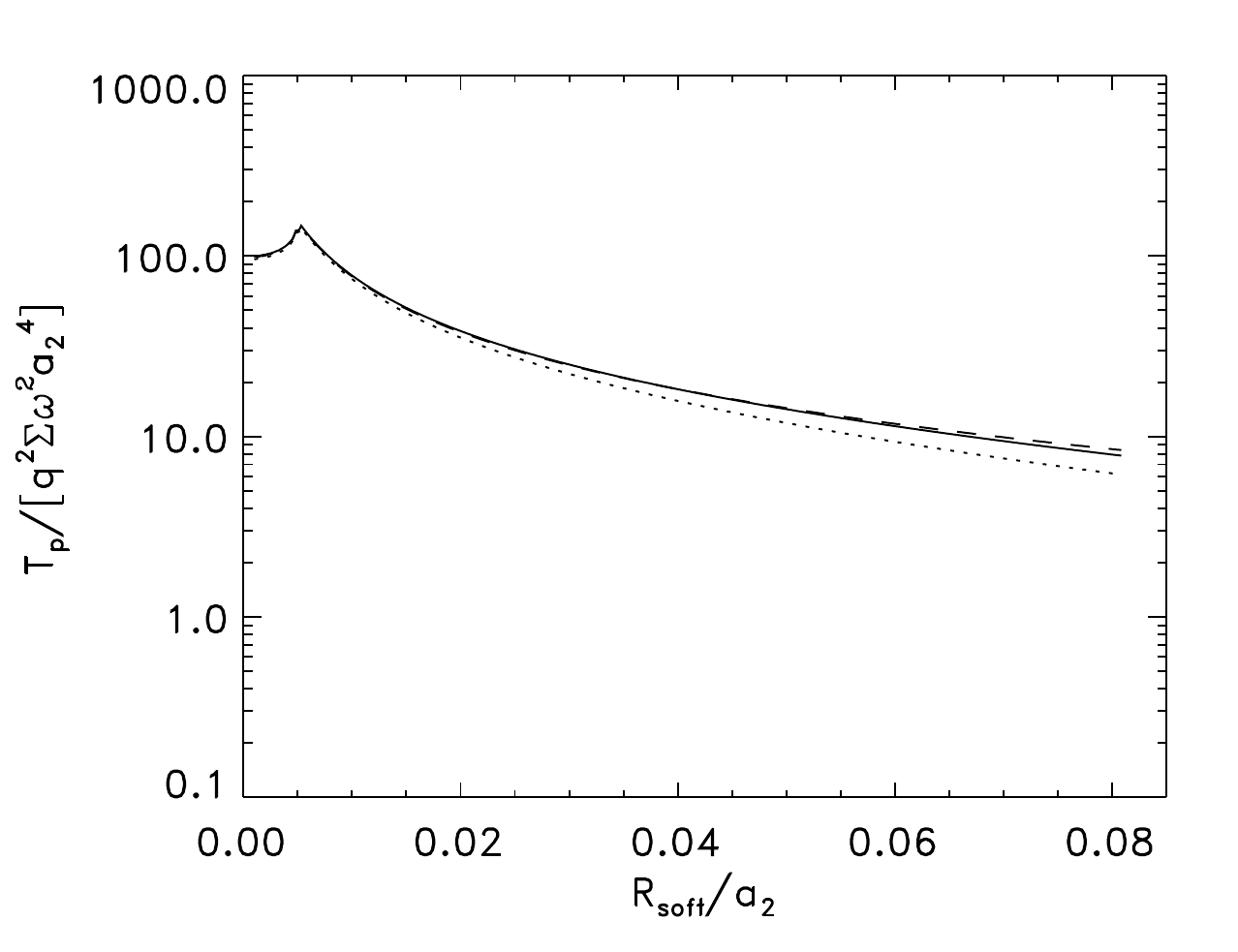}
  \caption{Dimensionless torque as a function of the softening radius for $q=10^{-3}$
using the orbital scattering approach. The solid curve corresponds to the torque derived
using Equations (\ref{eq:torqueKK}) and (\ref{eq:tgdelta}) with $\Delta_{1}=\Delta_{2}=0.2a_{2}$.
The dotted and dashed lines are the torques derived using Equations (\ref{eq:Tg_orbital_scatter}) 
and (\ref{eq:tgdelta}) with $\Delta_{1}=\Delta_{2}=0.2a_{2}$ (dotted line) and with
$\Delta_{1}=\Delta_{2}=0.5a_{2}$ (dashed line).
 }
\vskip 0.75cm
\label{fig:torque_scattering_q}
\end{figure}

The net torque acting on the {\rm perturber}, $T_{p}$, can be calculated by integrating $dT_{g}/dR$
over $R$ or, equivalently, over $\Delta$.
Since our perturbation analysis is strictly valid only for $\xi \ll a_{2}$,
which implies $\Delta \ll a_{2}$, we compute the net torque as
\begin{equation}
T_{p}= -\int_{-\Delta_{1}}^{\Delta_{2}}
\frac{dT_{g}}{d\Delta}\, d\Delta,
\label{eq:tgdelta}
\end{equation}
where the integral limits $\Delta_{1}$ and $\Delta_{2}$ must be taken less than $a_{2}$.
Note that the integration in Equation (\ref{eq:tgdelta}) diverges
if $R_{\rm soft}=0$ and $\Delta \rightarrow 0$. The reason is that 
Equation (\ref{eq:Tg_orbital_scatter}) is not valid for small
impact parameters if $R_{\rm soft}=0$. In fact, in the integration over $\Delta$, we have to exclude
the streamlines that suffer significant deflection. The condition for small
deflections is given in Eq. (\ref{eq:condition1}) in the Appendix.
For a point mass ($R_{\rm soft}=0$),
it requires that the impact parameter $\Delta$ should be larger than
$20R_{\rm acc}$.

Figure \ref{fig:torque_scattering_q} shows $T_{p}$ as a function of $R_{\rm soft}$ for $q=10^{-3}$. We compute
$T_{p}$ using Eqs. (\ref{eq:torqueKK}) and (\ref{eq:tgdelta}),
for $\Delta_{1}=\Delta_{2}=0.2a_{2}$. 
For comparison we also plot the net torque derived using Equation (\ref{eq:Tg_orbital_scatter}) instead of Equation
(\ref{eq:torqueKK}), for $\Delta_{1}=\Delta_{2}=0.2a_{2}$ and for $\Delta_{1}=\Delta_{2}=0.5a_{2}$.  
We see that the three 
curves almost overlap. The curves have an inflection point at 
$R_{\rm soft}\simeq 0.005a_{2}$. For values of $R_{\rm soft}$ smaller than $0.005a_{2}$,
$T_{p}$ flattens because of the exclusion of those impact parameters that produce large deflections.

The scattering approach is usually adopted to estimate the torque density in the nonlinear 
type II regime. For perturbers with $q$ small enough that $R_{\rm acc}\ll R_{\rm soft}$, 
the response of the disk is linear even in the immediate vicinity of the perturber and thus
the torque $T_{p}$ can be computed using linear theory.
\citet{mut11} derived the linear dynamical friction force on a softened object moving in a
rectilinear orbit with constant velocity $V_{\rm rel}$ through a 2D (zero thickness) 
slab of constant surface density $\Sigma$ and sound speed $c_{s}$.
For a supersonic perturber, the dynamical friction force, according to 
\citet{mut11}, is
\begin{equation}
F_{\rm \scriptscriptstyle DF}\simeq \frac{\pi \Sigma G^{2}M_{2}^{2}}{R_{\rm soft}V_{\rm rel}^{2}}.
\end{equation}
Applying this formula to a body on a retrograde circular orbit and
making use of the relations $GM_{2}=q\omega^{2}a_{2}^{3}$ and
$V_{\rm rel}=2\omega a_{2}$, the torque is
\begin{equation}
T_{p}=a_{2}F_{\rm \scriptscriptstyle DF}=\frac{\pi}{4} q^{2} \Sigma \omega^{2} a_{2}^{4} 
\left(\frac{a_{2}}{R_{\rm soft}}\right).
\label{eq:muto}
\end{equation}
In Figure \ref{fig:torque_comparison}, we see that the curve for $T_{p}$ derived in linear 
theory closely follows the curve derived in the scattering calculation.

As an alternative method, \citet{iva15} study the linear interaction between a 
retrograde circular-orbit satellite (secondary) and a 2D disk,
by assuming that the flow is stationary in a frame rotating with the secondary.
The torque $T_{p}$ is written as a summation of the contribution of the different 
Fourier modes with azimuthal wave number $m$. According to \citet{iva15}, 
the torque in the linear regime is
\begin{eqnarray}
&&T_{p}=2\pi q^{2} \Sigma \omega^{2}a_{2}^{4}\left(\frac{ a_{2}}{H}\right) \nonumber\\
&&\times \sum_{m=1}^{\infty}
\frac{m}{\sqrt{4m^{2}-1}} \exp\left(-2\sqrt{4m^{2}-1} R_{\rm soft}/H\right),
\label{eq:ivanov_lintorque}
\end{eqnarray}
where $H\equiv c_{s}/\omega$ with $c_{s}$ the sound speed evaluated at $R=a_{2}$.
Equation (\ref{eq:ivanov_lintorque}) can be approximated by
\begin{equation}
\begin{aligned}
T_{p}& \simeq  \frac{2\pi}{\sqrt{3}} q^{2} \Sigma \omega^{2}a_{2}^{4}\left(\frac{a_{2}}{H}\right) \\
&\times \biggl(\exp\left[-\frac{2\sqrt{3}R_{\rm soft}}{H}\right]+\frac{2}{\sqrt{5}}
\exp\left[-\frac{2\sqrt{5}R_{\rm soft}}{H}\right] \\
&+\frac{\sqrt{3}}{2} \frac{\exp(-12R_{\rm soft}/H)}{1-\exp(-4R_{\rm soft}/H)}\biggl).
\label{eq:ivanov_approx}
\end{aligned}
\end{equation}
If $R_{\rm soft}\ll 0.2H$, the last term in the right-hand side of Equation (\ref{eq:ivanov_approx}) 
is the leading term,
and the expression (\ref{eq:muto}) for the torque is recovered.  Indeed, it easy to check that Equation 
(\ref{eq:muto}) is a good
approximation of Equation (\ref{eq:ivanov_approx}) for $R_{\rm soft}\leq 0.3H$ with a fractional
error less than $16\%$.
In this limit, the torque $T_{p}$ is essentially independent of $H$ (see Eq. \ref{eq:muto}).

\begin{figure}
\includegraphics[width=88mm,height=65mm]{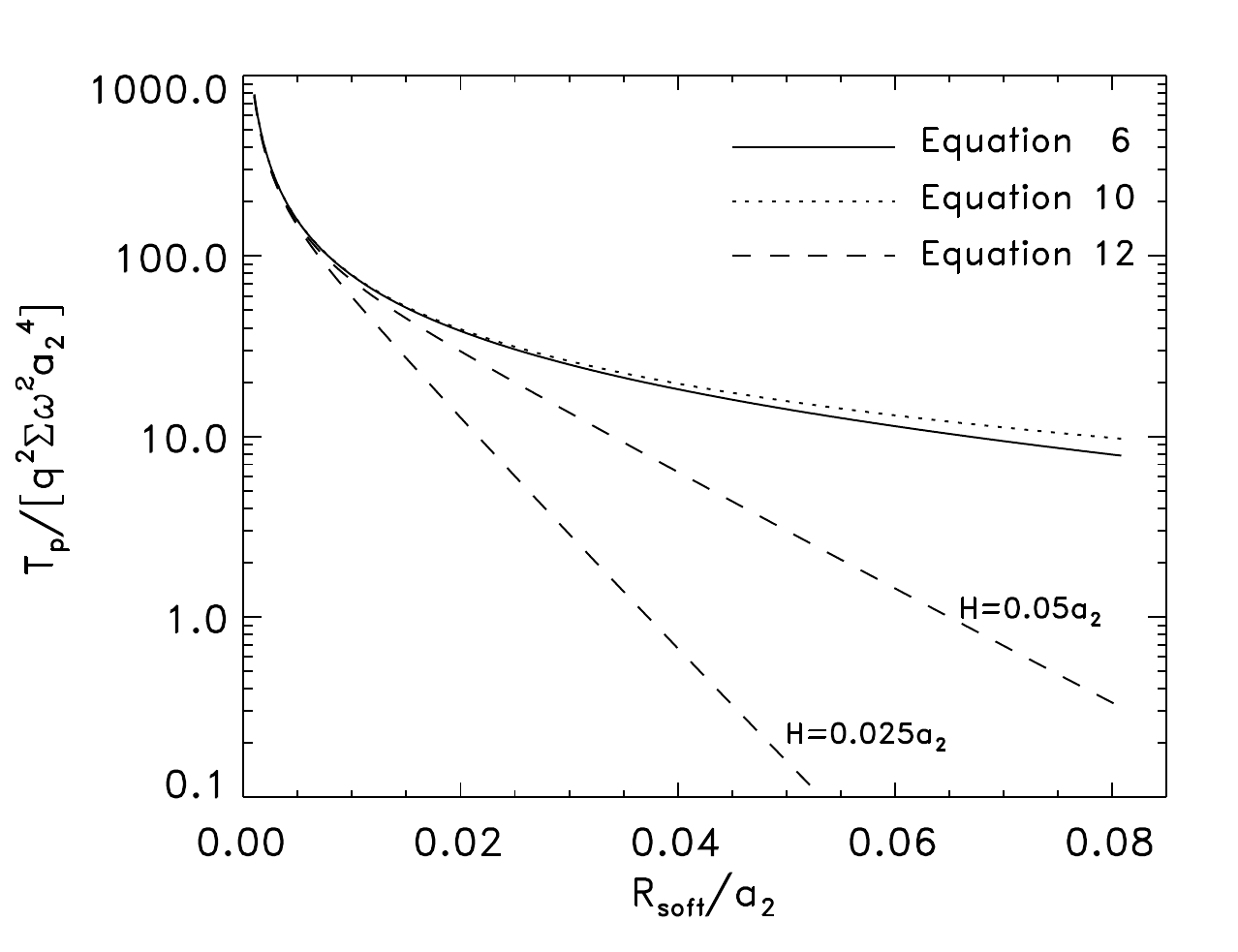}
  \caption{Dimensionless torque as a function of the softening radius
using different approaches. The solid line is calculated after adopting 
Equations (\ref{eq:torqueKK}) and (\ref{eq:tgdelta}) with $\Delta_{1}=\Delta_{2}=0.2a_{2}$,
the dotted lines using Equation (\ref{eq:muto}), and the dashed lines correspond to
Equation (\ref{eq:ivanov_approx}).
In all cases it is assumed that $q$ is small enough that $R_{\rm acc}\ll R_{\rm soft}$.
 }
\vskip 0.75cm
\label{fig:torque_comparison}
\end{figure}

For $R_{\rm soft}>0.3H$, the third term in the right-hand side of
Equation (\ref{eq:ivanov_approx}) is small compared to the two first terms. Given $H$, 
Equation (\ref{eq:ivanov_approx}) then predicts that
the torque declines exponentially when increasing $R_{\rm soft}$ with a characteristic scale $\sim H/4$.
On the other hand, if $R_{\rm soft}$ is kept fixed, 
Equation (\ref{eq:ivanov_approx}) predicts that the magnitude of the torque is sensitive
to the value of $H$. In particular, if $H\rightarrow 0$ (i.e. $c_{s}\rightarrow 0$), then $T_{p}\rightarrow 0$.

The above features are exemplified in Figure \ref{fig:torque_comparison} where we compare $T_{p}$
versus $R_{\rm soft}$, under the different approaches.
The values of the torque predicted by Equation 
(\ref{eq:ivanov_approx}) converge to the values derived in the scattering 
method {\it only} for small values of $R_{\rm soft}/H$. 
For values $R_{\rm soft}/H\gtrsim 1$,
Equation (\ref{eq:ivanov_approx}) predicts smaller torques than the other two methods.
For instance, for $R_{\rm soft}=0.04a_{2}$ and $H=0.025a_{2}$ (implying $R_{\rm soft}=1.6H$), 
Equation (\ref{eq:ivanov_approx})
predicts a torque about $30$ times lower than Equations (\ref{eq:torqueKK}) or (\ref{eq:muto}).  

The estimates of $T_{p}$ given in Eqs. (\ref{eq:torqueKK}) and (\ref{eq:muto}) ignore the 
gravitational torque exerted by the perturbed disk at the front side of the satellite. 
If we drop the satellite at $t=0$ in an unperturbed disk, these estimates should
be good at least at times $t\leq \tau_{1/2}$, that is, before the 
satellite has reached its own wake. At $t\gg \tau_{1/2}$, Equation (\ref{eq:ivanov_approx})
should be more precise as it takes into account the contribution of this
material. Equations (\ref{eq:torqueKK}), (\ref{eq:muto}) and (\ref{eq:ivanov_approx}) 
converge if $R_{\rm soft}\ll H$. 
The reason for this matching at large $H$ is that in thicker disks, the wake is less colimated 
when the perturber penetrates into it and hence the gravitational push is less important.
In fact, the condition $R_{\rm soft}\ll H$ implies that the thickness of the wake,
when the perturber reaches its tail (which is $c_{s}\tau_{1/2}$)
is much larger than $R_{\rm soft}$.

In Section \ref{sec:hydro_sims}, we present the results of 2D simulations
in order to answer the following questions: Does the torque take a value as predicted by 
Equations (\ref{eq:torqueKK}) and (\ref{eq:muto}) at $t\simeq \tau_{\scriptscriptstyle 1/2}$? 
At later times,
does the torque decrease with time to the value predicted by Equation (\ref{eq:ivanov_approx})? 
If so, what is the timescale for this relaxation process?

\subsection{Torques in a retrograde 3D disk}
\label{sec:3D_torques}

The 2D analysis is strictly valid only if the vertical velocity
dispersion of the disk particles is zero and thus the disk
has zero thickness at any time. 2D models correctly 
describe the far field (i.e. at distances from the perturber much larger than $H$),
but a delicate vertical averaging procedure is
necessary to obtain the 2D equations to account for the otherwise
neglected vertical thickness of the disk \citep[e.g.,][]{mul12,kle12}.

In the context of type I migration, it was soon noticed that the 2D
theory could give a reasonable estimate of the 3D torque in the {\it prograde} case,  
even if the perturber has a size much less than $H$ (e.g., \S 5.3 in Artymowicz 1993; Takeuchi
\& Miyama 1998). 
This is because of the existence of the torque cut-off:
acoustic waves cannot be excited at a distance from the perturber less than $\sim H$
if the perturber is on a prograde orbit
because the gas velocity relative to the perturber is subsonic and, in a stationary
state, a subsonic perturber cannot excite acoustic waves. 
More quantitatively, \citet{tan02} computed the torque in linear theory for a 2D
disk and for a 3D disk. In the particular case of a disk with
constant surface density, the torque for the 3D case is only $18\%$ higher than it is for the 
2D case.

However, if the secondary has a {\it retrograde} orbit, its relative velocity with respect to the gas is
supersonic at any distance from the perturber. As a consequence, the perturber can excite acoustic
waves at distances $\leq H$, especially if it is a point-mass with $R_{\rm acc}\ll H$. 
These waves carry angular momentum and contribute to the torque.
Therefore, a 3D treatment of the interaction is required to evaluate
the torque. The 3D analysis is also important if we wish to capture
3D effects in 2D simulations.

\subsubsection{Analytical predictions}

\citet{can13} use the ballistic approximation to compute the dynamical 
friction force on a gravitational perturber moving along a rectilinear trajectory, at relative 
velocity with respect to the gas $V_{\rm rel}$, in the midplane ($z=0$) 
of a vertically-stratified slab with density $\rho(z)=\rho_{0}\exp(-z^{2}/2H^{2})$.

For a perfect accretor ($R_{\rm soft}=0$) with $R_{\rm acc}\ll H$, 
\citet{can13} obtain that the drag force is
\begin{equation}
F_{\rm \scriptscriptstyle DF}=\frac{\sqrt{8\pi}\Sigma (GM_{2})^{2}}{V_{\rm rel}^{2} H} 
\ln \left(\left[\frac{8e^{3}}{\pi} \right]^{1/2} \frac{H}{R_{\rm acc}}\right),
\label{eq:canto_BH}
\end{equation}
where $\Sigma$ is the surface density of the slab ($\Sigma=\sqrt{2\pi}\rho_{0}H$). 
Using again that for a perturber in counter-rotating circular orbit $V_{\rm rel}=2\omega a_{2}$ and
$GM_{2}=q\omega^{2} a_{2}^{3}$,
the corresponding torque is
\begin{equation}
T_{\rm bal}^{(r)}=\sqrt{\frac{\pi}{2}} \frac{q^{2}}{h} \Sigma \omega^{2} a_{2}^{4}
\ln \left( \frac{14.3 h}{q}\right),
\label{eq:torque_canto_accretor}
\end{equation}
where we have used that $R_{\rm acc}=qa_{2}/2$. 
It is important to remark here that the aerodynamic torque due to
mass accretion by the perturber, which is given by
\begin{equation}
T_{\rm acc}^{(r)}=\frac{\pi}{2} \frac{q^{2}}{h}\Sigma \omega^{2} a_{2}^{4},
\label{eq:accretion_torque}
\end{equation}
has also been included in $T_{\rm bal}^{(r)}$. The aerodynamic torque
$T_{\rm acc}^{(r)}$ is not negligible; for $h/q=50$, $T_{\rm acc}^{(r)}$ contributes
$19\%$ to the total torque $T_{\rm bal}^{(r)}$.

For a non-accreting perturber with softening radius much smaller than $H$, \citet{can13}
infer that 
\begin{equation}
F_{\rm \scriptscriptstyle DF}=\frac{\sqrt{8\pi}\Sigma (GM_{2})^{2}}{V_{\rm rel}^{2} H} 
\ln \left(\frac{r_{\rm max}}{r_{\rm min}}\right),
\end{equation}
where $r_{\rm max}\simeq 2.1\sqrt{2}H$ and $r_{\rm min}$ is the minimum impact parameter
for the interaction. In the case of a Plummer perturber with $R_{\rm soft}>R_{\rm acc}$, 
\citet{ber13}
found that $r_{\rm min}=2.25R_{\rm soft}$ \citep[see also][]{san99}.
Putting together, the predicted torque on a softened body having $R_{\rm acc}<R_{\rm soft}\ll H$,
on a retrograde circular orbit, is
\begin{equation}
T_{\rm bal}^{(r)}=\sqrt{\frac{\pi}{2}} \frac{q^{2}}{h} \Sigma \omega^{2} a_{2}^{4}
\ln \left(\frac{1.32 H}{R_{\rm soft}}\right).
\label{eq:torque_canto_softened}
\end{equation}

On the other hand, \citet{iva15} also give an expression for the torque
on a small perturber ($R_{\rm acc}<H$ and $R_{\rm soft}<H$) using the linear
mode decomposition. If the disturbing body is
not very massive (more specifically, for $q$ less than $3\pi h/5$, which is the 
range of interest in the present investigation), they find 
\begin{equation}
T_{\rm wave}^{(r)}= \frac{5}{3} \frac{q^{2}}{h} \Sigma \omega^{2} a_{2}^{4},
\label{eq:ivanov_3d}
\end{equation}
without distinction between accreting or non-accreting perturbers. 
Equation (\ref{eq:ivanov_3d}) is similar to Equations (\ref{eq:torque_canto_accretor}) 
and (\ref{eq:torque_canto_softened}), except by
factors of order unity and the logarithmic dependence. 
By comparing Eq. (\ref{eq:accretion_torque}) and (\ref{eq:ivanov_3d}), we find
that $T_{\rm wave}^{(r)}$ is only $6\%$ larger than $T_{\rm acc}^{(r)}$,
which suggests that $T_{\rm wave}^{(r)}$ might underestimate the net torque on an 
accreting perturber. Moreover, $T_{\rm bal}^{(r)}$
may be significantly larger than $T_{\rm wave}^{(r)}$.
For instance, for a perfect accretor ($R_{\rm soft}=0$) with $h/q$ between
$50$ and $500$, $T_{\rm bal}^{(r)}$ is a factor of $5-7$ larger than $T_{\rm wave}^{(r)}$.
In the next Section, we present the results of 3D simulations to
test the accuracy of these two formulae.

\section{Hydrodynamical simulations}
\label{sec:hydro_sims}

We have carried out 2D and 3D simulations of a binary embedded in a gaseous disk, 
using the FARGO3D code \citep{ben16}\footnote{FARGO3D is a publicly available code at http://fargo.in2p3.fr.}. 
We use a spherical grid $(r,\phi,\theta)$. The self-gravity of the disk is not included.
In all our simulations, the secondary with a mass $q=10^{-3}$ is dropped 
suddently at $t=0$ and it is held on a fixed circular orbit in retrograde direction.
Initially, the disk has pure azimuthal rotation and constant surface density.
The thermal profile is fixed by assuming that
the ratio between the sound speed and the circular velocity is constant with $R$.
The evolution is locally isothermal. The kinematical viscosity of the disk was taken
$\nu=10^{-5} \omega a_{2}^{2}$ in all the simulations. We simulate a ring of the disk
with inner radius $r_{\rm min}$ and outer radius $r_{\rm max}$. 
In the radial direction we employ damping boundary conditions at $r_{\rm min}$
and at $r_{\rm max}$ \citep{deV06}. 
Remind that the disk rotates in counterclockwise direction and the binary in
clockwise direction: the secondary loses angular momentum if the torque 
$T_{p}$ is positive.

Two dimensionless parameters define a model: $h\equiv H/a_{2}=c_{s}/(\omega a_{2})$ 
and $s\equiv R_{\rm soft}/a_{2}$. 
In terms of $h$, a retrograde perturber in circular orbit travels at a Mach number 
${\mathcal M}=2/h$ relative to the local gas.

We have checked that convergence of the results requires at least $2$ zones per
$R_{\rm soft}$ in the radial direction, and $1.5$ zones per $R_{\rm soft}$
in the azimuthal and vertical directions.

\subsection{2D simulations}
\label{sec:2Dsims}
In all the simulations presented in this subsection, we use $r_{\rm min}=0.3a_{2}$
and $r_{\rm max}=4a_{2}$, and the number of zones per $R_{\rm soft}$ is taken to be larger than
$2$ in each direction, in order to obtain reliable estimates of the torque.
We have explored the effect of the finite size of the domain by comparing models at 
varying size of the domain (see Figure \ref{fig:torque_boundaries}).
In all these models, the number of zones per $R_{\rm soft}$
is $9$ in the radial direction and $1.5$ in the azimuthal direction and, therefore, all
they have the same spatial resolution. We find that for 
$r_{\rm min}\lesssim 0.3a_{2}$ and $r_{\rm max}\gtrsim 3a_{2}$, the torque
is estimated with reasonable accuracy. In fact, the torque for the model having $r_{\rm min}=0.2a_{2}$
and $r_{\rm max}=6a_{2}$ (not shown) overlaps with the curve obtained for 
$r_{\rm min}=0.25a_{2}$ and $r_{\rm max}=3.5a_{2}$.

\begin{figure}
\hbox{\hspace{-1.7em}\includegraphics[width=95mm,height=70mm]{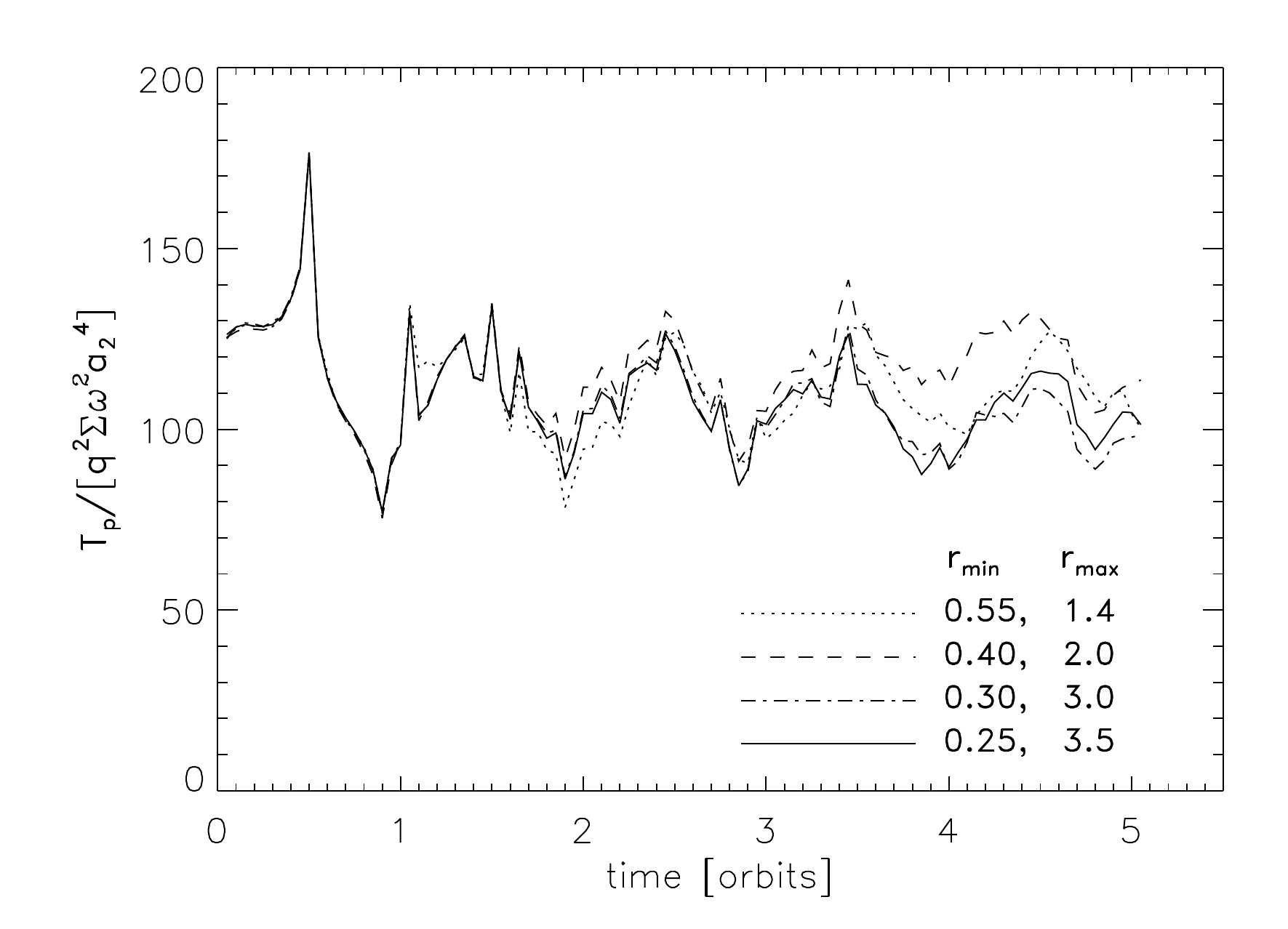}}
  \caption{Dimensionless torque versus time, in 2D runs, for different values of $r_{\rm min}$ and
$r_{\rm max}$ (in units of $a_{2}$), but having the same number of zones per $R_{\rm soft}$. 
In all cases, $h=0.06$ and $s=0.006$. 
 }
\vskip 0.75cm
\label{fig:torque_boundaries}
\end{figure}

\begin{figure*}
\center
\hbox{\hspace{0.9em}\includegraphics[width=150mm,height=125mm]{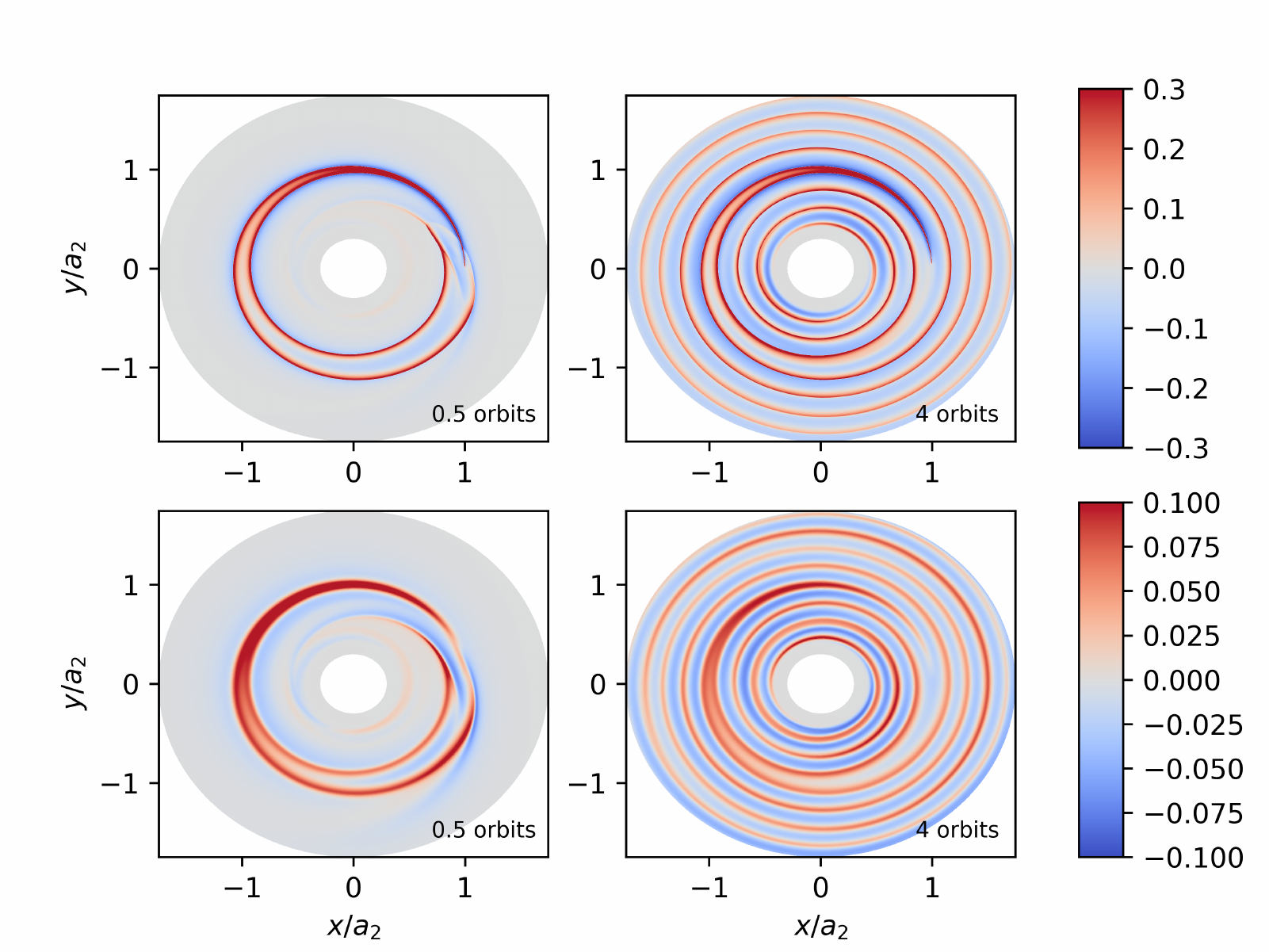}}
  \caption{Snapshots of the perturbation of the surface density $(\Sigma-\Sigma_{i})/\Sigma_{i}$, where
$\Sigma_{i}$ is the initial surface density, in 2D runs with $h=0.05$ and $s=0.006$ (upper panels),
and with $h=0.05$ and $s=0.06$ (lower panels). The time is quoted at the lower-right corner in
each panel. The secondary is at $x/a_{2}=1$ and $y=0$ in all the panels. 
Only the central regions of the disk is shown (the computional domain extends up to $r_{\rm max}=4a_{2}$).
 }
\vskip 0.75cm
\label{fig:maps}
\end{figure*}

\
\begin{figure}
\hbox{\hspace{-1.1em}\includegraphics[width=92mm,height=165mm]{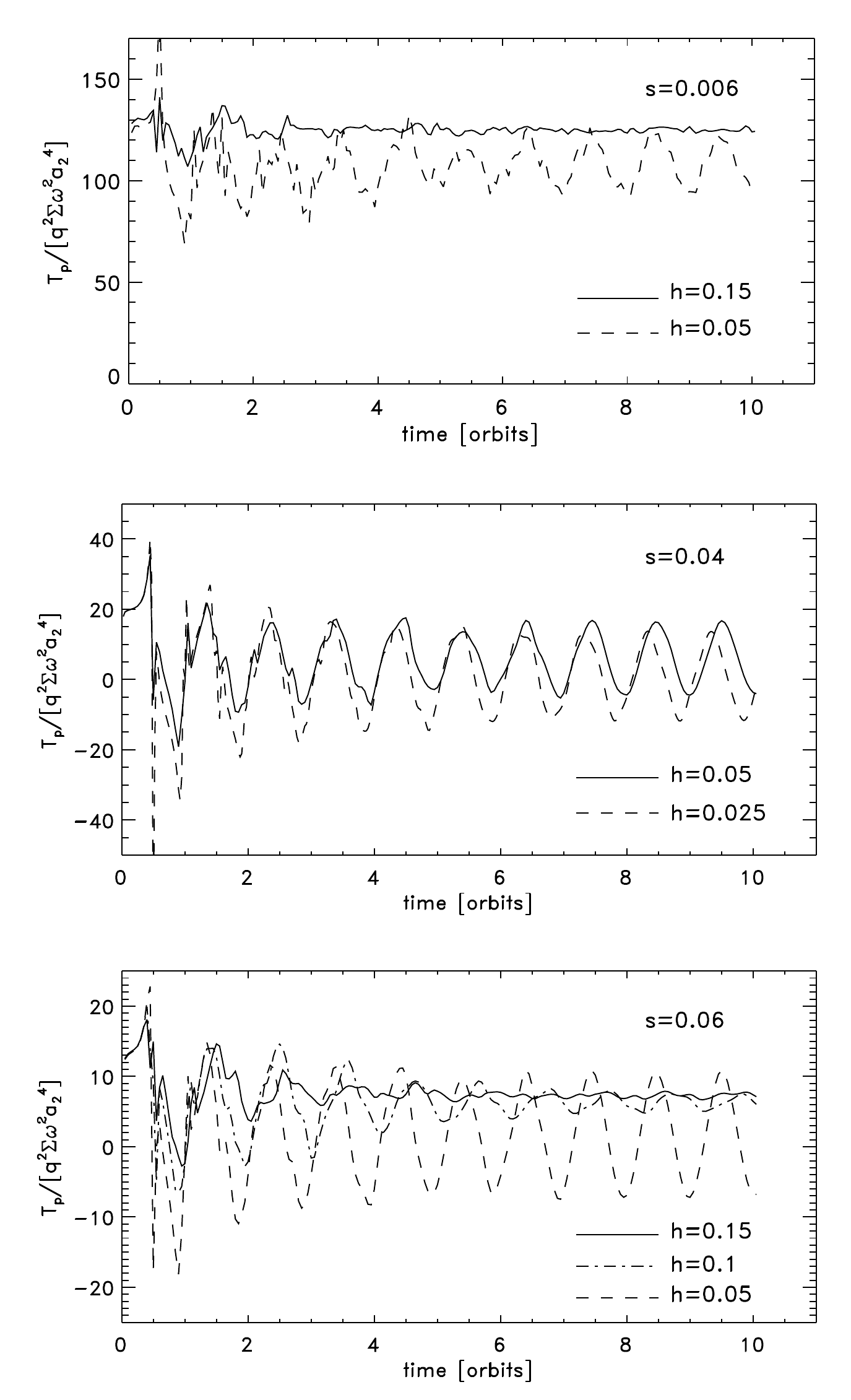}}
  \caption{Dimensionless torque as a function of time in 2D simulations with 
different combinations of $h$ and $s$ (see legends).  
}
\vskip 0.75cm
\label{fig:torque_time}
\end{figure}

Our first objective is to find out the dependence of the strength of the torque with $R_{\rm soft}$
and with the gas sound speed (or, equivalently with $h$).
To do so, we have computed the torque on the perturber for several values 
of $s$, spanning between $0.006$ and
$0.06$, and different values of $h$, between $0.025$ and $0.15$. 

Figure \ref{fig:maps} displays color maps of the perturbed surface density 
for a disk with $h=0.05$ and two different softening radii
($s=0.006$ and $s=0.06$). The density wake consists of very thin plumes that are curved 
along the orbit. These plumes are bounded by sharp discontinuities.
After a few orbits, the density wake becomes tighly wound.

\begin{figure}
\center
\hbox{\hspace{-1.1em}\includegraphics[width=92mm,height=165mm]{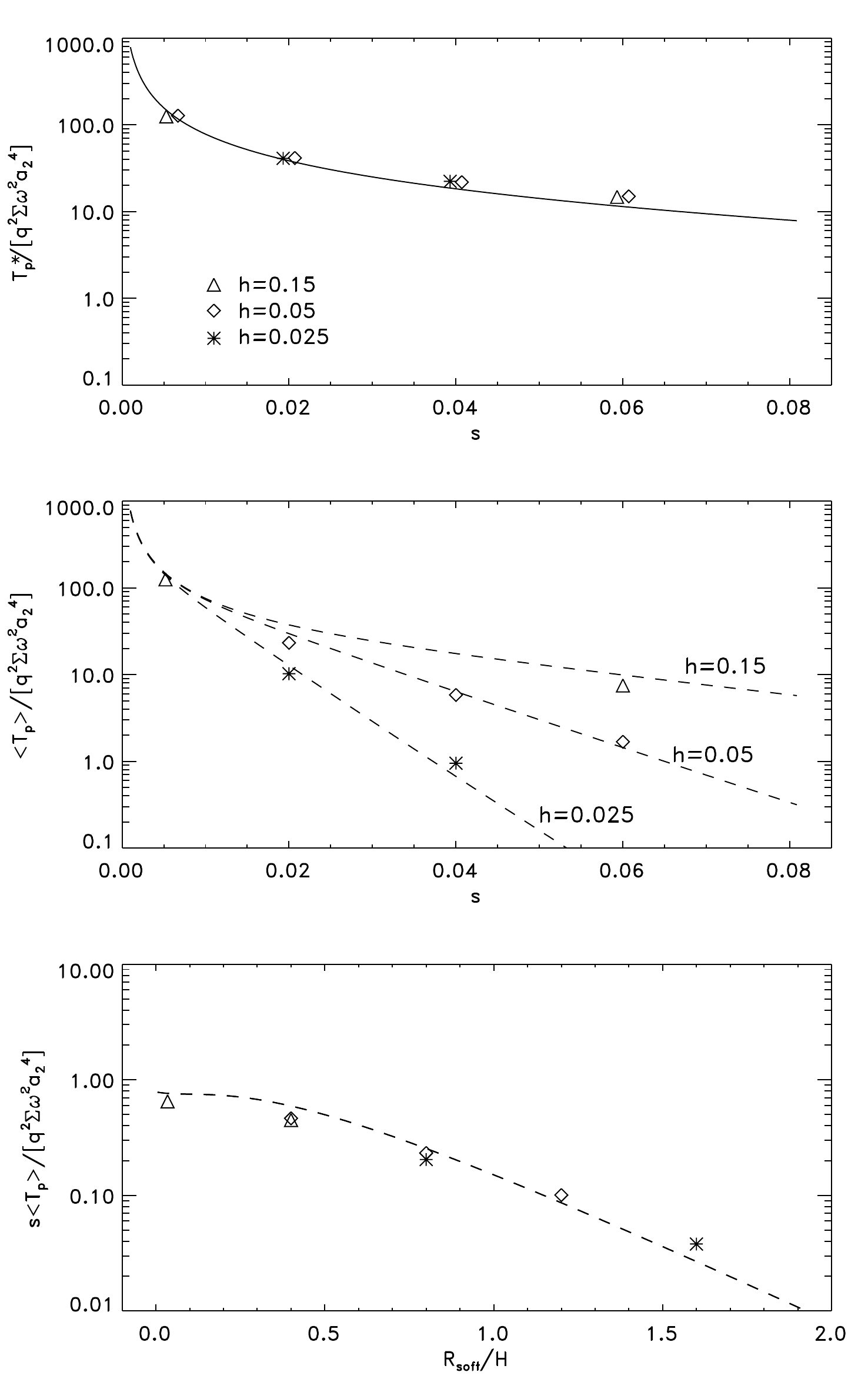}}
  \caption{Comparison of the torque in 2D simulations (symbols) and analytical
predictions (lines). The top panel shows $T_{p}^{\ast}$ for different
simulated models, while the solid line indicates the torque using the ballistic
approximation presented in \S \ref{sec:excitation_torque}. 
A small shift along the horizontal axis has been introduced to avoid overlapping.
The middle panel shows 
$\left<T_{p}\right>$ versus $s$ for the same models. The dashed
lines represent the curves using Equation (\ref{eq:ivanov_approx}), for varying values
of $h$. The bottom panel shows $s\left<T_{p}\right>$ as a function of $R_{\rm soft}/H$.
The dashed line corresponds to the curve predicted by Equation (\ref{eq:ivanov_approx}).
}
\vskip 0.75cm
\label{fig:torque_models_sim}
\end{figure}

The behaviour of the torque versus time is shown in Figure \ref{fig:torque_time}
for seven representative cases. Consider first the magnitude of the torque at early stages, i.e.
at $t<0.4$ orbits, when the perturber has not penetrated yet into its own wake.
We see in this Figure that the torque at those times depends on $s$ 
but it is not sensitive to the disk aspect ratio $h$, i.e. it does not
depend on the sound speed of the gas. This is in agreement
with Equations (\ref{eq:torqueKK}) and (\ref{eq:muto}).
The reason is that the perturber moves so highly supersonic that
the trajectory of a fluid element can be obtained by neglecting
the pressure force. In this limit, the sound speed of the gas is not
relevant in determining the torque.
This pressureless approximation was already made by \citet{bon44} to study mass 
accretion onto a star. In order to check further
whether or not the ballistic approximation developed in \S \ref{sec:excitation_torque} reproduces
satisfactorily  the torque at early times, it is convenient to define $T_{p}^{\ast}$ as the mean torque 
between $t=0.05$ and $t=0.4$ orbits. The upper panel of
Figure \ref{fig:torque_models_sim} compares $T_{p}^{\ast}$ as obtained from our simulationes with 
the torque predicted by Equations (\ref{eq:torqueKK}) and (\ref{eq:tgdelta}). We find a good 
agreement between them.

At $t\simeq 0.5$ orbits, the secondary encounters its own tail. At this time, in most of the cases,
the torque exhibits a remarkable peak and then an abrupt decline.  
This is the result of the interaction with small-scale density substructures in the wake, as
the secondary runs close to them (see Figure \ref{fig:maps}).
Indeed, during the first $2.5$ orbits, the disk is at the process of relaxation and
the torque may exhibit some narrow fluctuations until those density substructures are 
smeared out. Note, however, that these fluctuations in the torque are small for disks with
$h= 0.15$.

After $t\simeq 2.5$ orbits, the torque reaches a nearby constant value in
the models with $h=0.15$. In these models, 
the wake achieves a quasi-stationary state in the frame rotating with the perturber. 
For a compact body (i.e. $s=0.006$),
and a relatively hot disk ($h=0.15$), its stationary value is almost identical
to $T_{p}^{\ast}$ (see first panel of Figure \ref{fig:torque_time}).

On the other hand, in models with $h\leq 0.05$, the torque oscillates around
the mean value with frequency $\omega$. To be sure that these oscillations are not a
numeric artefact, we varied the viscosity, the value of the mass of the secondary,
and used different rotating frames of reference, as well as different initial radial profiles for 
the disk surface density.
In all these variations, the oscillations in the torque are present. At a fixed value of $c_{s}$, say at
$h=0.05$, the amplitude of these oscillations is rather insensitive to the adopted
value of $s$. All this suggests that disks with $h\leq 0.05$ are trapped in an
oscillatory mode. A long-time run shows that these oscillations are persistent 
(see Figure \ref{fig:torque_longtime}); they are damped in a timescale $\sim 30$ orbits 
orbits but they can be again self-excited in the disk.

The torque in the model with $h=0.1$ and $s=0.06$ 
(see third panel in Figure \ref{fig:torque_time}), presents oscillations, but they are 
damped after $\sim 5$ orbits.

The orbital decay timescale of the secondary is of course determined by the average torque  
over time. Figure \ref{fig:torque_models_sim} shows $\left<T_{p}\right>$, which denotes the mean 
torque between $t=3$ and $t=10$ orbits (i.e. it is the average taken over $7$ complete cycles). For 
comparison, we also plot the predicted values according to Equation (\ref{eq:ivanov_approx}). We find
a good agreement between them. 

\begin{figure}
\hbox{\hspace{-1.1em}\includegraphics[width=92mm,height=68mm]{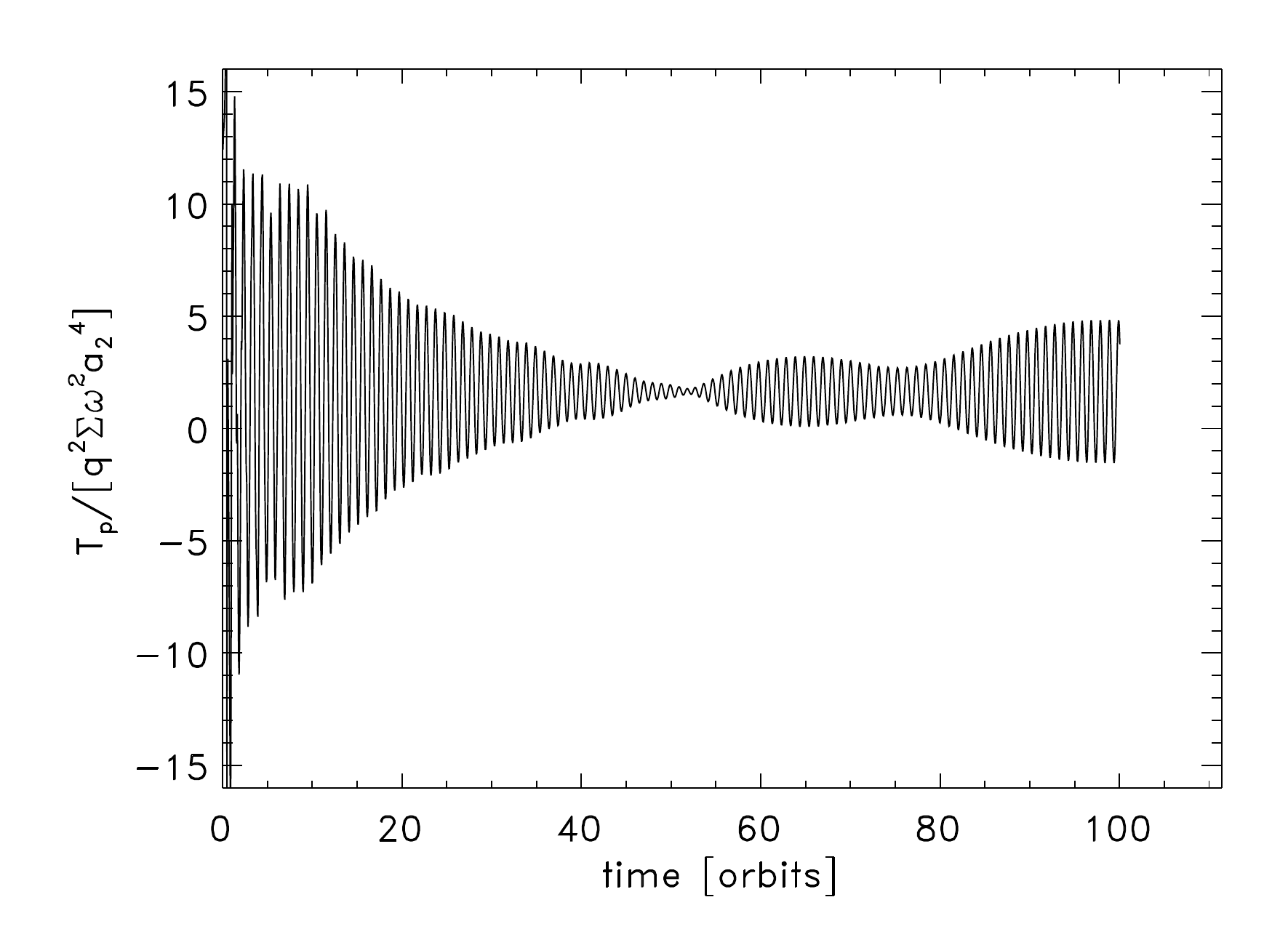}}
  \caption{Dimensionless torque as a function of time in a 2D run with $h=0.05$ and
 $s=0.06$.
 }
\vskip 0.75cm
\label{fig:torque_longtime}
\end{figure}

\begin{table*}
	\centering
	\caption{Parameters of the 3D runs}
\label{table:params} 
 \begin{tabular}{|c|c|c|c|c|c|c|c|}\hline
 Run &  $h$ &  $s$ & $R_{\rm soft}/H$ & $r_{\rm min}$ & $r_{\rm max}$ & $z_{\rm max}$ & 
$(N_{r},N_{\phi},N_{\theta})$ \\ 
               
\hline 
 $1$       &  $0.06$   &  $0.003$  &  $0.05$  &  $0.55a_{2}$ & $1.4a_{2}$ & $2.6H$ & $(512, 2560, 96)$ \\
$2$   &  $0.06$   &  $0.003$  &  $0.05$  &  $0.55a_{2}$ & $1.4a_{2}$ & $2.6H$ & $(1024, 5120, 192)$ \\ 
$3$   &  $0.15$   &  $0.0075$  &  $0.05$  &  $0.3a_{2}$ & $3.5a_{2}$ & $2.7H$ & $(1560, 2048, 192)$   \\ 
$4$   &  $0.18$   &  $0.018$  &  $0.1$  &  $0.65a_{2}$ & $1.3a_{2}$ & $1.7H$ & $(1536, 1536, 768)$  \\ 

 \hline 
\end{tabular}  
\end{table*}

To assess the extent to which the initial radial profile may affect the torque, we repeated
the simulations starting with $\Sigma\propto R^{-3/2}$. For the range of $h$ and $s$ under 
consideration, the maximum change in the mean torque $\left<T_{p}\right>$ with respect
to a flat disk, occurs for the model with 
$s=0.06$ and $h=0.025$, for which the torque differs by $30\%$. On the other hand,
for models with $s=0.006$, $\left<T_{p}\right>$ for $\Sigma\propto R^{-3/2}$ is almost
identical to the torque for $\Sigma=$const.

In summary, the formula of \citet{iva15} for the torque of a satellite in retrograde 
circular orbit predicts correctly the mean torque. The formula derived in the scattering
approach gives the correct value of the torque for compact perturbers ($R_{\rm soft}\leq 0.3H$).
It also provides the strength of the torque in the first half orbit, i.e. before the
perturber reaches its own wake. 
In the next subsection we extend the numerical
analysis to small-size gravitational objects embedded in full 3D disks.

\begin{figure}
\hbox{\hspace{-0.7em}\includegraphics[width=88mm,height=65mm]{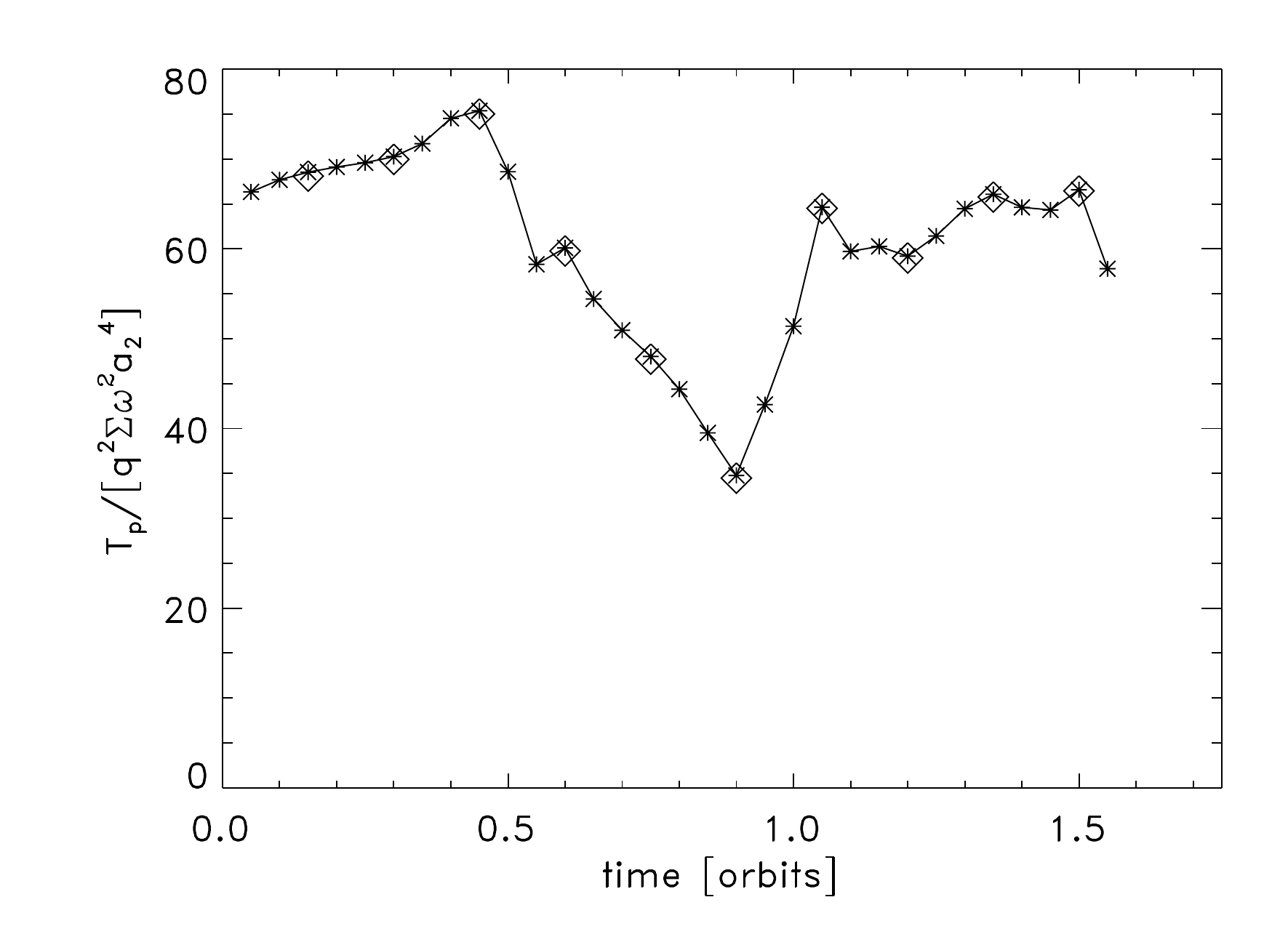}}
  \caption{Dimensionless torque versus time for Run 1 and Run 2, which
differ only in the resolution (see Table \ref{table:params}). 
The diamonds correspond to Run 1, whereas the asterisks correspond to Run 2.
Both simulations have $h=0.06$ and $s=0.003$ (or, equivalently, $R_{\rm soft}=H/20$).
}
\vskip 0.75cm
\label{fig:torque_5120}
\end{figure}

\subsection{3D simulations}

We wish to determine the 3D structure of the wake created  
by a retrograde perturber with a physical size much smaller than the local scaleheight
of the disk $H$. 
We run a set of $4$ simulations of a retrograde perturber with $q=10^{-3}$ in a 
fixed circular and coplanar orbit. Table \ref{table:params} lists the parameters of 
the simulations: $h$, $s$, $R_{\rm soft}/H$, the computational domain and the resolution
used. $z_{\rm max}$ denotes the vertical extension of our box at the perturber's location.
In the four simulations, the spacing between grid points is $<R_{\rm soft}$.
Note that for $R_{\rm soft}<H/10$ (or, equivalently, for $s<h/10$), we have
$T_{\rm bal}^{(r)}/T_{\rm wave}^{(r)}>2$. Therefore, models with $s< h/10$ may be used to
discern between Equation (\ref{eq:torque_canto_softened}) and Equation (\ref{eq:ivanov_3d}).

Figure \ref{fig:torque_5120} shows the torque versus time over the first $1.5$ orbits in Run 1 
($h=0.06$ and $s=0.003$).
Doubling the number of zones in each direction (Run 2) gives the same values for the torque 
(differences less than $1\%$). As occurs in the corresponding 2D run, the torque exhibits a first
peak at $t=0.45$ orbits (although less pronounced than it is in the 2D counterpart) and a minimum 
at $t=0.9$ orbits.

Figure \ref{fig:Tstar_3D} shows the values of $T_{p}^{\star}$ derived from our
simulations, together with the torque predicted in the scattering
approximation (Equation \ref{eq:torque_canto_softened}). A good agreement between numerical and
analytical values are found.

In order to determine the mean torque, it is preferable to consider cases where the torque
shows small fluctuations. Otherwise, we must run the simulation for a long time, what is 
computationally very expensive in 3D simulations. According to our 2D runs in 
Section \ref{sec:2Dsims}, this can be accomplished by using a thicker disk.

Figure \ref{fig:torque_h015} shows the temporal evolution of the torque for $h=0.15$ and $s=0.0075$ 
(again $R_{\rm soft}=H/20$). For comparison, we also plot $T_{\rm bal}^{(r)}$
(Equation \ref{eq:torque_canto_softened}) and $T_{\rm wave}^{(r)}$ 
(Equation \ref{eq:ivanov_3d}). It is clear that $T_{\rm bal}^{(r)}$ predicts correctly the
mean value of the torque in this case, while $T_{\rm wave}^{(r)}$ underestimates the torque.

In summary, we have confirmed that $T_{\rm bal}^{(r)}$
reproduces correctly the value of the torque acting on a retrograde 
object with $R_{\rm soft}$ significantly smaller than $H$.
For systems with $R_{\rm soft}\gtrsim 0.3H$, $T_{\rm bal}^{(r)}$ overestimates
the torque, because it ignores the gravitational pull of the wake ahead of the perturber.

\begin{figure}
\hbox{\hspace{-1.7em}\includegraphics[width=95mm,height=70mm]{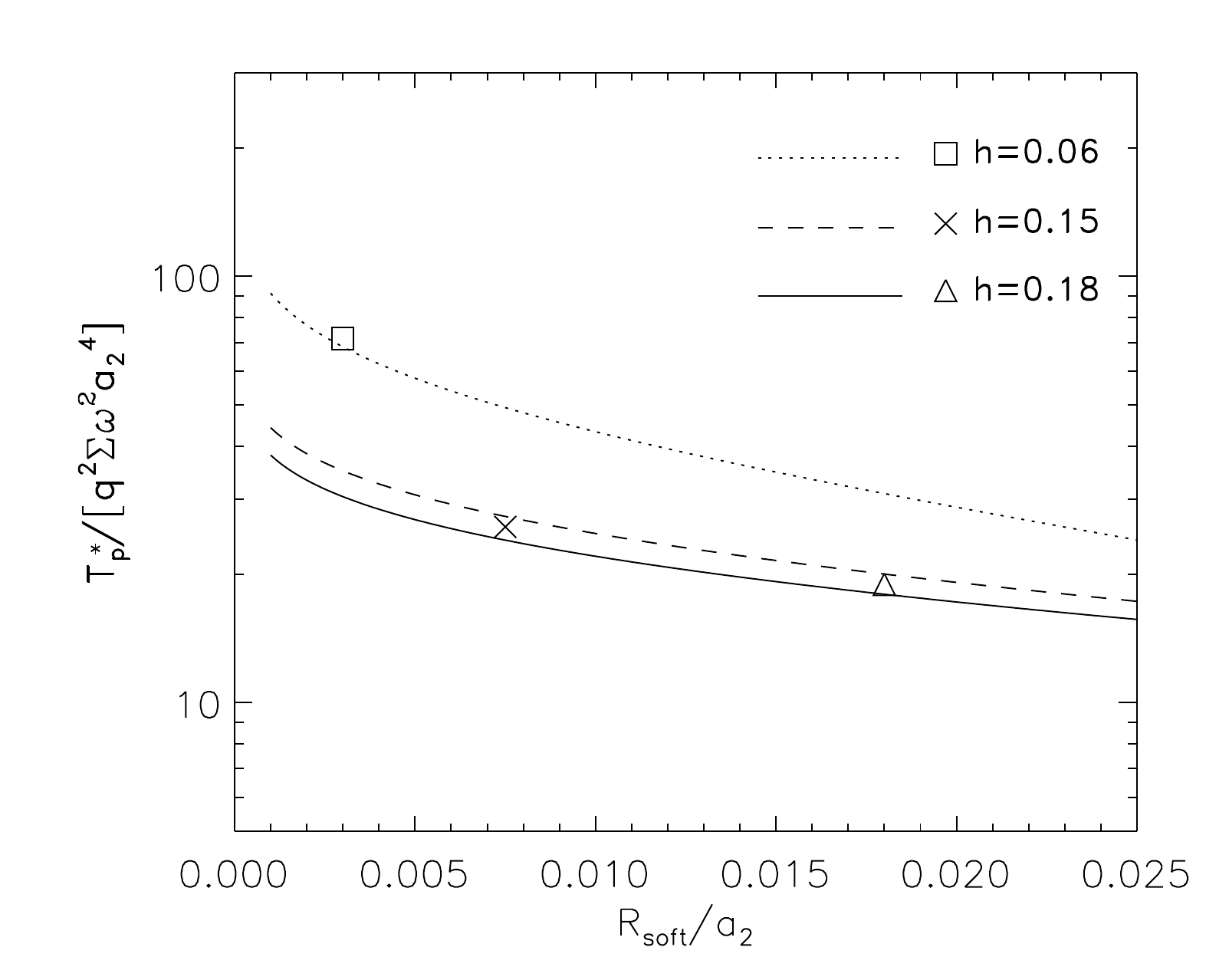}}
  \caption{$T_{p}^{\ast}$ is plotted for 3D simulations with different combinations of $h$ and $s$  
(symbols). This torque is compared with Equation (\ref{eq:torque_canto_softened}).
}
\vskip 0.75cm
\label{fig:Tstar_3D}
\end{figure}

\begin{figure}
\hbox{\hspace{-1.7em}\includegraphics[width=95mm,height=70mm]{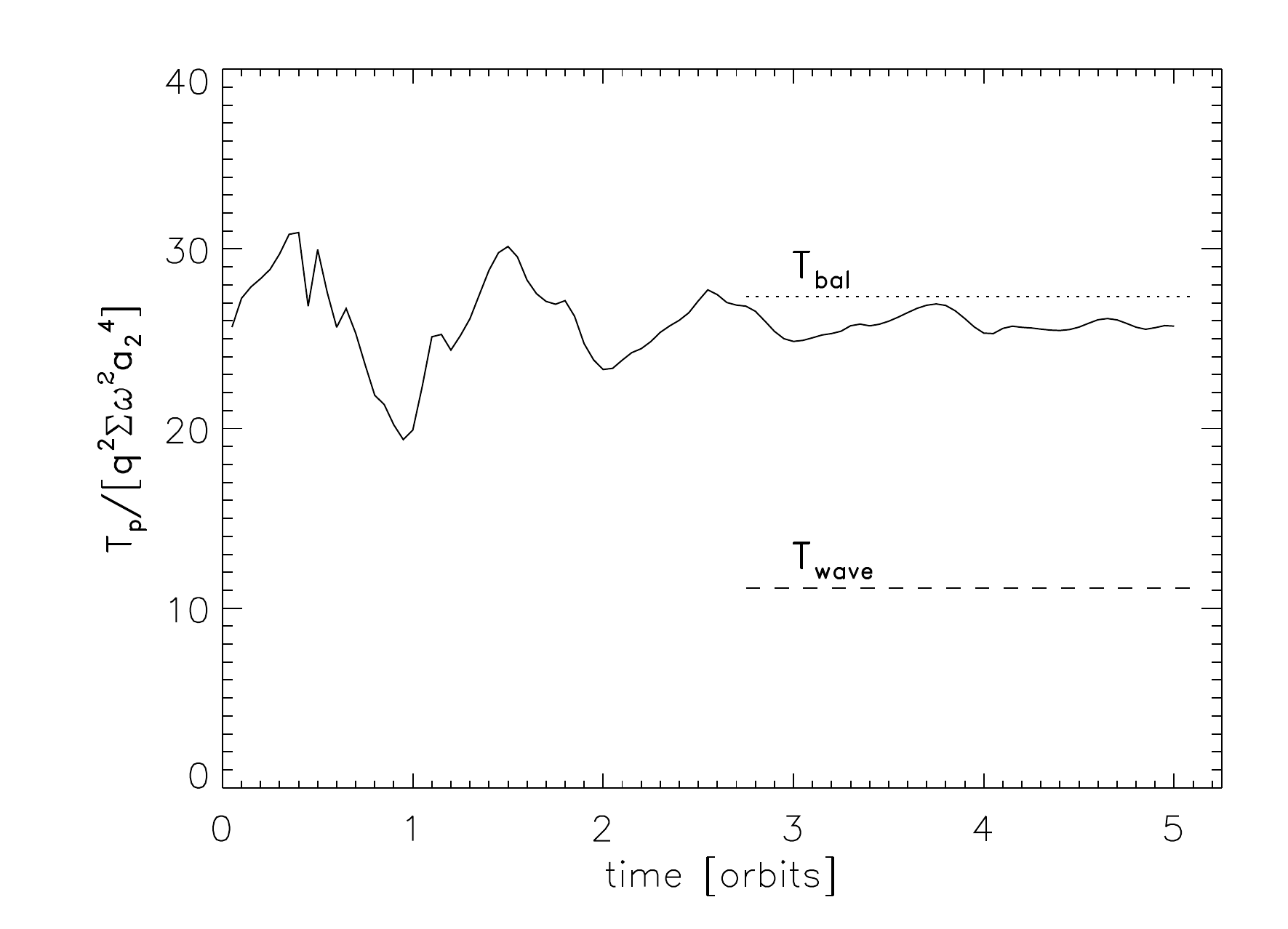}}
  \caption{Dimensionless torque as a function of time for Run 3 ($h=0.15$ and $s=0.0075$).
The torque is compared
with the values predicted by the ballistic formula $T_{\rm bal}^{(r)}$
(Eq. \ref{eq:torque_canto_softened}) and with the torque $T_{\rm wave}^{(r)}$ (Eq. \ref{eq:ivanov_3d}).
 }
\vskip 0.75cm
\label{fig:torque_h015}
\end{figure}

\section{Discussion}
\label{sec:discussion}

Perturbers with $q\leq q_{\rm cr}^{(r)}$ satisfy $R_{\rm acc}\ll H$ if they
are embedded in a gaseous disk with aspect ratio and viscosity expected in
astrophysical disks (Section \ref{sec:assumptions}). 
Thus, we may use Equation (\ref{eq:torque_canto_accretor})
to estimate the torque acting on a retrograde BH binary (or on a binary
composed by a massive BH and a star) with $q\leq q_{\rm cr}^{(r)}$. 
Remind that Equation (\ref{eq:torque_canto_accretor}) includes the 
contribution of the non-linear part of the wake close to the body, as well as mass accretion.
In the following, we discuss some implications of our results.

\subsection{Comparing migration timescales for BH binaries: prograde versus retrograde disks}
\label{sec:timescales_comparison}
For a satellite with semimajor axis $a_{2}$, the characteristic timescale
for radial migration is defined as
\begin{equation}
\tau\equiv \frac{a_{2}}{\dot{a}_{2}}= \frac{M_{2}\omega a_{2}^{2}}{2|T_{p}|}.
\label{eq:migration_timescale_gen}
\end{equation}

It is usually argued that the radial migration of low-mass
companions in coplanar retrograde motion is much slower than in standard prograde
migration scenarios \citep[e.g.,][]{mck14,ban15,iva15}.
In the following, we compare radial migration
timescales for binaries in prograde and in retrograde orbits, assuming orbits
are always in the midplane of the disk.

Using Equation (\ref{eq:torque_canto_accretor}) for the torque, the resulting type I 
migration timescale is
\begin{equation}
\tau_{\rm \scriptscriptstyle I}^{(r)}=\sqrt{\frac{\pi}{2}} \frac{h}{q\eta\omega \ln(14.3h/q)},
\end{equation}
where $\eta$ is the ratio between the characteristic disk mass contained within the orbital radius
$\pi \Sigma_{2} a_{2}^{2}$ and the central mass $M_{1}$, i.e. $\eta\equiv \pi \Sigma_{2}a_{2}^{2}/M_{1}$.
Here $\Sigma_{2}$ is the disk surface density at the position of the secondary\footnote{\citet{iva99}
also estimated a characteristic timescale for the orbit to change but in the case that the the orbit
is inclined by an angle larger than $h$, resulting in a larger timescale by a factor of $\sim h^{-1}$
\citep[see also][]{xiang13}.}.

Consider now a prograde binary with $q\ll 3h^{3}\leq q_{\rm cr}^{(p)}$. 
In this case, the response of the disk is linear \citep[e.g.,][]{miy99}.
The torque acting on the secondary was evaluated by \citet{tan02}
for a 3D disk with surface density $\Sigma \propto R^{-\beta}$. They find that the
torque for type I migration is
\begin{equation}
T_{\scriptscriptstyle \rm I}^{(p)}=\psi(\beta) \frac{q^{2}}{h^{2}}\Sigma_{2}\omega^{2}a_{2}^{4},
\label{eq:TIp}
\end{equation}
where $\psi (\beta)=1.364+0.541\beta$. 
Once the torque is known, we may evaluate the type I migration timescale in the prograde
case, $\tau_{\rm \scriptscriptstyle I}^{(p)}$, from Eq. (\ref{eq:migration_timescale_gen}).

The dashed line in Figure \ref{fig:migration_time} gives $\tau_{\rm \scriptscriptstyle I}^{(p)}$
versus $q$ for a disk with $\beta=0$ and $h=0.05$, while the solid line 
corresponds to $\tau_{\rm \scriptscriptstyle I}^{(r)}$.
The ratio between the migration timescale for a
BH in retrograde orbit and prograde orbit is
\begin{equation}
\frac{\tau_{\rm \scriptscriptstyle I}^{(r)}}{\tau_{\rm \scriptscriptstyle I}^{(p)}}= 
\frac{1.088+0.432\beta}{h\ln(14.3h/q)}.
\end{equation}
For $\beta=0$ and $h=0.05$, this
ratio is $2.45$ for a binary with $q=10^{-4}$. However, for disks
thicker than $h\simeq 0.1$, the radial migration rate may be larger (shorter migration timescale) in the 
retrograde case than in the prograde.

Now consider values of $q$ in the range $q_{\rm cr}^{(p)}<q<q_{\rm cr}^{(r)}$, where $q_{\rm cr}^{(p)}$ 
and $q_{\rm cr}^{(r)}$ are
given in Equations (\ref{eq:qcr_prograde}) and (\ref{eq:qcr_ac}), respectively.
As guide numbers, we have $q_{\rm cr}^{(p)}=3\times 10^{-4}$ and 
$q_{\rm cr}^{(r)}=2.4\times 10^{-3}$ for a disk with parameters $h=0.05$ and $\alpha=3\times 10^{-3}$.
For mass ratios between $q_{\rm cr}^{(p)}$ and $q_{\rm cr}^{(r)}$, 
the secondary component has enough mass to carve a gap in a prograde disk, and will
undergo type II migration with timescale $\tau_{\rm \scriptscriptstyle II}^{(p)}$. However, the 
secondary weakly disturbs the disk if it is in retrograde orbit (see \S \ref{sec:assumptions}),
and migrates in a timescale $\tau_{\rm \scriptscriptstyle I}^{(r)}$. In the following, we compare
$\tau_{\rm \scriptscriptstyle II}^{(p)}$ with $\tau_{\rm \scriptscriptstyle I}^{(r)}$.

In the standard descriptions of prograde type II migration,
the timescale is estimated as a first approximation by
\begin{equation}
\tau_{\rm \scriptscriptstyle II}^{(p)}= 
\frac{2}{3\alpha h^{2}\omega}\times {\rm max}\left(1,\frac{\pi q}{4\eta}\right)
\end{equation}
(Sotiriadis et al. 2017, and references therein). 
However, more detailed simulations have demonstrated that the type II migration
timescale in the prograde case can be larger or smaller than $\tau_{\rm \scriptscriptstyle II}^{(p)}$, 
depending on the disk mass \citep{duf14,dur15}.

In Figure \ref{fig:migration_time} we plot the migration timescales as a function of $q$
for an accretor embedded in a disk 
with parameters $h=0.05$, $\eta=0.01$ and $\alpha$ between $3\times
10^{-3}$ and $10^{-2}$. We also plot $\tau_{\rm acc}^{(r)}\equiv
M_{2}\omega a_{2}^{2}/(2T_{\rm acc}^{(r)})$. 
We see that for $\alpha=3\times 10^{-3}$, the migration timescale for a retrograde
accretor with $q=10^{-3}$ is $\sim 90$ times 
shorter than the viscous drift timescale $\tau_{\rm \scriptscriptstyle II}^{(p)}$, and 
about $\sim 20-30$ times shorter than the corresponding migration timescale found in 
prograde simulations by \citet{duf14} and \citet{dur15}.

It is remarkable that for $q$'s in the range $q_{\rm cr}^{(p)}<q<q_{\rm cr}^{(r)}$ and for
the values of the viscosity under consideration,
$\tau_{\rm acc}^{(r)}$ is smaller by a factor of $2-10$ than 
$\tau_{\rm \scriptscriptstyle II}^{(p)}$ (see Figure \ref{fig:migration_time}).
In other words, the accretion torque on a satellite in retrograde rotation is larger than the net torque
in prograde rotation.

\begin{figure}
\includegraphics[width=88mm,height=65mm]{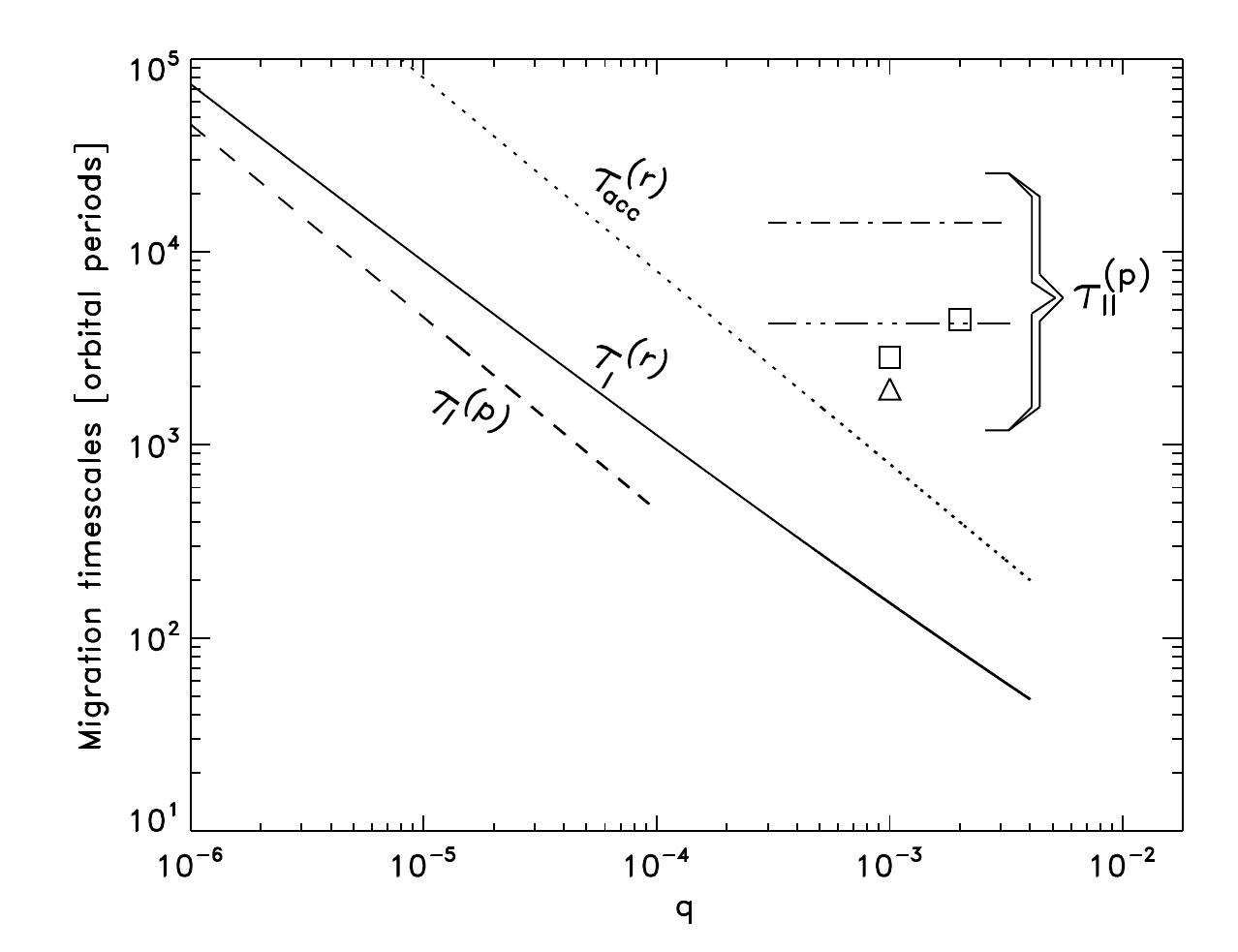}
  \caption{Comparison of the migration timescales for circular binaries with different ratios 
of secondary to primary mass $q$, for a disk with $\beta=0$, $h=0.05$ and $\eta=0.01$.
The solid line represents the migration timescale for retrograde binaries.
The dashed line corresponds to the type I migration timescale for prograde
binaries. This regime is adequate provided that $q\lesssim 10^{-4}$.
The horizontal dot-dashed lines correspond to the viscous drift timescale
for $\alpha=0.003$ (upper line) and for $\alpha=0.01$ (lower line).
We also plot the migration timescale for prograde binaries in the type II
regime from the numerical calculations by \citet{dur15}, who assume
$\alpha=0.003$ (squares), and from \citet{duf14}, who use $\alpha=0.01$
(triangle).
}
\vskip 0.75cm
\label{fig:migration_time}
\end{figure}

\subsection{The effective smoothing length in retrograde 2D simulations}

In numerical simulations, point-like satellites, such as planets and BHs, are usually described
with softened Plummer gravitational potentials to avoid singularities in the evaluation of the force.
In 3D simulations, the smoothing length is chosen as small as possible while meeting convergence
constraints. However, the smoothing length in 2D runs is usually chosen to resemble
a stratified 3D disk as closely as possible \citep{mas02,mul12}.
One route is to choose such a smoothing length in 2D simulations that the
total torque acting on the perturbing particle is approximately equal to that obtained
in 3D simulations. For prograde disks, 2D and 3D torques are similar for 
smoothing lengths $R_{\rm sl}$ in the range $(0.5-0.7)H$
\citep{tan02,mas02,mul12}.
Likewise, the required amount of
smoothing in retrograde 2D runs can be estimated by equating the 2D torque given 
in Equation (\ref{eq:ivanov_approx}), with the 3D torque given in Equation (\ref{eq:torque_canto_accretor})
\begin{eqnarray}
&&\sqrt{\frac{8\pi}{3}}\exp(-2\sqrt{3}\lambda)
+\frac{2}{\sqrt{5}}\exp(-2\sqrt{5}\lambda)\\ \nonumber
&&+\frac{\sqrt{3}}{2} \frac{\exp(-12\lambda)}{1-\exp(-4\lambda)} =\ln (14.3h/q),
\end{eqnarray}
where $\lambda\equiv R_{\rm sl}/H$, with $R_{\rm sl}$ the smoothing length
in 2D runs. We see that $\lambda$ depends on $h/q$. In particular, for $h/q=50$, we obtain
$R_{\rm sl}=0.095H$, whereas for $h/q=500$, we get $R_{\rm sl}=0.068H$.
$R_{\rm sl}$ is significantly smaller than $H$ because the 
torque contribution of the wake at distances $\sim H$ from the perturber is not negligible. 
We must stress, however, that these values for $R_{\rm sl}$ do not ensure that the surface
density in 2D runs will match the projected surface density in full 3D disks.

\subsection{The process of alignment between the disk and the BH orbital plane}
In \S \ref{sec:timescales_comparison}, we have discussed the radial
migration timescale under the assumption that the angular momentum of the binary 
$\vecJ_{b}$ is antiparallel to the angular momentum of the disk $\vecJ_{d}$ (i.e. orbital
inclination $i=180$ deg).
This counter-aligned orientation is stable if $J_{d}<2J_{b}$ \citep{kin05,lod06,nix12}.
If this condition is not fulfilled,
torques can flip the binary orbit, and the system evolves towards the prograde state
of rotation, even if the binary starts out close to counter-alignment
(i.e. the binary evolves towards prograde coplanarity ($i=0$), passing first
through a polar orbit).

The stability condition for a circular binary with $q\ll 1$ can be
written as $M_{d}/M_{1} \lesssim 2q\sqrt{a_{2}/R_{d}}$, where $R_{d}$ is a characteristic
radius for the disk whose mass is $M_{d}$. We have taken $J_{d}\simeq M_{d}\sqrt{GM_{1}R_{d}}$
and $J_{b}=M_{2}\sqrt{GM_{1}a_{2}}$. For a typical disk with $M_{d}\simeq 10^{-2}M_{1}$
and a secondary with $q\lesssim 10^{-3}$, it is clear that the criterion
is not satisfied.

The temporal evolution of the orbital parameters of an inclined perturber
due to successive interactions with a disk has been examined in different
works \citep{sye91,rau95,vok98,rei12}.
All these studies assume that the force acting on the perturber, at the moment
that the perturber crosses the disk, is antiparallel 
to the velocity of the perturber relative to the disk, which is a good
approximation also in our case. For circular orbits with inclinations $|\sin i|
\gg h$, ${\mathcal{R}}\equiv a_{2}\cos^{4}(i/2)$ is a constant, i.e.
$d{\mathcal{R}}/dt=0$ along the orbit evolution \citep{rau95,vok98}.
From the constancy of ${\mathcal{R}}$, 
the timescale for the inclination to change  can be related to the timescale
for radial migration. For instance,
a body that starts with an inclination of $165$ deg will reach one tenth
its initial orbital radius when it has an inclination of $153$ deg.
This implies that the timescale for variations in inclination is significantly 
larger than it is for radial evolution. However, we caution that the timescale
$a_{2}/\dot{a}_{2}$ may be sensitive to the inclination, especially at inclinations 
close to $180$ deg. In fact, a satellite with an inclination of $170$ deg
in a disk with $h=0.05$,
only spends about $25\%$ of its time at $|z|\leq H$. Therefore, precise estimates of the 
radial migration timescale for bodies in highly retrograde orbits should take into account the 
departure from coplanarity.

\section{Conclusions}
\label{sec:conclusions}

We have studied the gravitational coupling between a counter-rotating
satellite of mass $M_{2}$ and a disk. For simplicity, we have assumed that
the orbit of the perturber is circular and coplanar to the disk. 
In this situation, the perturber moves highly supersonic relative to the 
surrounding gas. 

We have evaluated the torque acting on a low-mass perturber in 2D hydrodynamical
simulations. In all the runs, the softening radius of the perturber is large enough
that the response of the disk is linear. Therefore, the
angular momentum imparted to the disk is not deposited locally but
is carried away by waves. We find that the formula derived in the ballistic
approximation can account for the strength
of the torque in the first half-orbit of the perturber, i.e. when the
perturber has not yet penetrated into its own wake. At later times, after the
perturber overtakes its own tail, the ballistic approximation predicts
correctly the strength of the torque as long as $R_{\rm soft}\leq 0.3 H$
(we remind that $H=c_{s}/\omega$, evaluated at the orbital radius $a_{2}$).
For larger values of the softening radius, the disk has not time to recircularize
between two successive passages of the perturbing mass. Therefore, the assumption
that disk particles return to a circular orbit breaks down. 

Our 2D runs indicate that the torque oscillates with time for perturbers
with $R_{\rm soft}>0.3H$. The amplitude of these oscillations may be large
for $R_{\rm soft}\gtrsim H$. For illustration, the amplitude of the oscillations
in the first $10$ orbits for a run with $h=0.05$ and $R_{\rm soft}=1.2H$, is a factor 
of $5$ times larger than the time-averaged torque. In this case, the oscillations in the
torque are damped after $\sim 30$ orbits.

We have studied how the time-averaged torque depends on the parameters
of the disk.
We find that the formula based on the Fourier mode decomposition carried out by 
Ivanov et al. (2015) predicts correctly the time-averaged torque in 2D runs. 
We confirm that the time-averaged torque is sensitive to the sound speed of the gas (i.e.
it depends on $H$) if $R_{\rm soft}>0.3H$.
For instance, if we take $R_{\rm soft}=0.06a_{2}$, the 2D torque (average over time) 
is a factor of $5$ larger when $H=0.15a_{2}$ than for $H=0.05a_{2}$.

It is worth noting that the time-averaged torque in 2D simulations
is sensitive to the adopted value for $R_{\rm soft}$.
For instance, for a disk with $h=0.05$, the torque is a factor of $5$
larger for $R_{\rm soft}=0.3H$ than for $R_{\rm soft}=0.7H$. This strong
dependence of the torque with $R_{\rm soft}$ should
be borne in mind when doing 2D simulations.

Real disks have finite thickness and, therefore, 2D calculations are an approximation to
the problem. The dependence
of the torque with the disk parameters (e.g., with $H$) is not necessarily 
the same for 2D disks than for 3D disks. A 3D treatment of the disk-perturber
interaction has been carried out in order to evaluate the torque exerted on the perturber 
by the near field, i.e. by the wake at distances $\leq H$. 

We have ran high-resolution, 3D hydrodynamical simulations of a
low-mass circular-orbit perturber interacting with a retrograde disk to confirm
the validity of the approximation of ballistic orbits. In our 3D
simulations, we take $R_{\rm soft}<0.1H$. The agreement
between the predicted torque in the ballistic approximation and the
numerical result is very good.

We have compared the radial migration timescale of prograde versus
retrograde perturbers. Provided that the mass ratio $q$ is so low
that the surface density of the disk is weakly disturbed, 
i.e.~for $q<{\rm min} (q_{\rm cr}^{(p)},q_{\rm cr}^{(r)})$, 
the migration timescale in the prograde case is shorter than in the retrograde
case if the aspect ratio of the disk is $\lesssim 0.1$. For thicker
disks, a body migrates faster if it rotates in the retrograde
sense than in the prograde sense.

Systems with $q$ in the interval $q_{\rm cr}^{(p)}<q<q_{\rm cr}^{(r)}$
can open a gap in the disk if they are in prograde orbit, yielding to
a saturation of the torque. However, this effect of ``decretion'' flow 
does not occur in the retrograde circular case.
As a consequence, the migration timescale of satellites in retrograde motion may be shorter 
than in prograde motion. For instance, for a disk with $h=0.05$ and
$\eta=0.01$, the migration timescale for a body in retrograde orbit is $20-30$ times 
shorter than in prograde orbit.
A related phenomenon occurs for $q>0.1$. As \citet{nix11} 
showed, accretion by the secondary in a retrograde disk is more efficient 
in shrinking a BH binary than accretion in the prograde case
because of the absence of a decretion flow in a retrograde binary.
Therefore, the statement that the torque is much weaker in retrograde disks
than in prograde disks because retrograde disks cannot support resonances
is not valid in general. 

Our results have also relevant implications regarding the issue of the 
appropriate smoothing length to be adopted in retrograde 2D runs.
We find that 2D runs require a smoothing length of $(0.07-0.1)H$ in order
to obtain the same torque than in 3D runs. Softening lengths just slighly
smaller than the disk scale height, as usually adopted in the prograde
case, underestimate the torque for retrograde perturbers.

\acknowledgments
We are grateful to the referee for very helpful comments which have improved the manuscript.
We thank Julio Clemente for technical support and Fr\'ed\'eric Masset for his advice and suggestions.
The 3D simulations were performed using MIZTLI Supercomputer of Universidad Nacional
Aut\'onoma de M\'exico (UNAM). The authors acknowledge financial support by PAPIIT project
IN111118.

\appendix
\section{Angular momentum exchange between a retrograde perturber and the disk}
\label{sec:appendix}
In this Appendix we determine the change in the orbital parameters
of a disk particle after a gravitational encounter with a perturber
of mass $M_{2}$ that is on a circular and retrograde orbit. 
We assume that in the absence of any perturber, all the particles of the disk move in
circular orbits, which are confined to the $z=0$ plane. Therefore, 
a disk particle with orbital radius $R_{0}$ rotates with
angular frequency $\Omega=\sqrt{GM_{1}/R_{0}^{3}}$, so that its azimuthal angle changes at
a rate $\dot{\phi}=\Omega$. In the presence of the mass $M_{2}$, the orbits of disk particles
will be perturbed in radius and azimuth. 
We consider a system of reference with origin at the center of the background potential $\Phi_{b}$, 
but ignore the indirect term in the potential \citep[see][for its definition]{nel00}
because of our assumption $q\ll 1$.
Let $(R(t),\phi(t))$ the radius and azimuthal angle of a disk particle.
We will treat $\Phi_{2}$ as a small perturbation
and write the perturbed orbit as $R(t)=R_{0}+\chi(t)$
and $\phi(t)=\Omega t+\alpha(t)$. The equations of motion are
\begin{equation}
\ddot{\chi}+(\kappa_{0}^{2}-4\Omega^{2})\chi-2\Omega R_{0}\dot{\alpha}=
-\frac{\partial \Phi_{2}}{\partial R}\biggl|_{0},
\label{eq:ddotR1}
\end{equation}
\begin{equation}
R_{0}\ddot{\alpha} +2\Omega\dot{\chi} =-\frac{1}{R_{0}}\frac{\partial \Phi_{2}}{\partial \phi}\biggl|_{0},
\label{eq:ddotalpha}
\end{equation}
where $\kappa_{0}$ is the radial epicyclic frequency at $R_{0}$
\citep[e.g.,][]{lin79,bin87}.
To derive $\chi(t)$ and $\alpha(t)$, we follow the same procedure as \citet{gol82},
which was also sketched in \citet{bin87}, Chapter 7.

The radial and azimuthal derivatives of $\Phi_{2}$ in the unperturbed orbit
are calculated using the local field approximation, which is valid when the duration of 
one gravitational encounter is $\ll \pi/\Omega$. This implicitly assumes that
$R_{\rm soft}\ll a_{2}$. In the retrograde case, we find
\begin{equation}
\frac{\partial \Phi_{2}}{\partial R}\biggl|_{0}=\frac{GM_{2}\Delta_{0}}{\xi_{0}^{3}(1+4B^{2} t^{2})^{3/2}},
\end{equation}
and
\begin{equation}
\frac{1}{R_{0}}\frac{\partial \Phi_{2}}{\partial \phi}\biggl|_{0}=\frac{2GM_{2}R_{0}\omega t}{\xi_{0}^{3}(1+4B^{2} t^{2})^{3/2}},
\label{eq:Phi_phi}
\end{equation}
where 
\begin{equation}
\Delta_{0}\equiv R_{0}-a_{2},\hskip  1.4cm
\xi_{0}\equiv (\Delta_{0}^{2}+R_{\rm soft}^{2})^{1/2}, \hskip 1.2cm {\rm and} \hskip 1.2cm
B\equiv \frac{\omega R_{0}}{(\Delta_{0}^{2}+R_{\rm soft}^{2})^{1/2}}.
\label{eq:BB}
\end{equation}
We recall that $\omega$ is the angular frequency of the mass $M_{2}$ around $M_{1}$.
Equations (\ref{eq:ddotR1})-(\ref{eq:BB}) 
describe the orbit of a disk particle, initially on a circular orbit,
during one single encounter with the perturber. 
For convenience we have chosen that the minimum distance between the disk particle and the
perturber occurs at $t=0$.

Upon substituting Equation (\ref{eq:Phi_phi}) into Equation (\ref{eq:ddotalpha}) yields 
\begin{equation}
R_{0}\dot{\alpha}+2\Omega \chi= -\frac{GM_{2}}{2\xi_{0}\omega R_{0} (1+4B^{2}t^{2})^{1/2}}.
\label{eq:dotalpha}
\end{equation}
Here we have imposed that $\dot{\alpha}=\chi=0$ at $t\rightarrow -\infty$.
Substituting $\dot{\alpha}$ from Equation (\ref{eq:dotalpha}) into Equation (\ref{eq:ddotR1}), 
we obtain
\begin{equation}
\ddot{\chi}+\kappa_{0}^{2}\chi=\frac{GM_{2}\Omega}{\xi_{0}^{2} B}
\left(1-\frac{\Delta_{0}}{\xi_{0}}\frac{B}{\Omega}+4B^{2}t^{2}\right)\left(1+4B^{2}t^{2}\right)^{-3/2}.
\end{equation}
Using the method of Fourier transform, we can find the asymptotic solution ($t\rightarrow \infty$)
for this differential equation $\chi(t)=A_{1} \sin \kappa_{0} t$ with
\begin{equation}
A_{1}=\left(\frac{GM_{2}\Omega}{\kappa_{0}\omega^{2}R_{0}^{2}}\right)
\bigg|-K_{0}\left[\frac{\kappa_{0}}{2B}\right]+
\frac{\Delta_{0}\kappa_{0}}{2\xi_{0}\Omega}K_{1}\left[\frac{\kappa_{0}}{2B}\right]\bigg|,
\label{eq:amplitude}
\end{equation}
where $K_{0}$ and $K_{1}$ are the modified Bessel functions.
Therefore, the orbit after the encounter can be described by an epicyclic motion with amplitude $A_{1}$
and eccentricity $e=A_{1}/R_{0}$.

\subsection{Keplerian disk}
In order to compute the torque between the disk and the disturbing mass $M_{2}$, we need to derive the change
of angular momentum of disk particles per unit of time. For a Keplerian disk, a disk particle initially 
has angular momentum per unit of mass $J_{0}=\sqrt{GM_{1}(a_{2}+\Delta_{0})}$. After one encounter
with the perturber, the angular momentum of the disk particle will be $J_{0}+\delta J$.
From the Jacobi integral, we may compute $\delta J$ as follows \citep[e.g.,][]{gol82}.
For a perturber in counter-rotation, conservation of Jacobi integral implies that
$E_{f}-E_{0}+\omega \delta J =0$, 
where $E_{0}$ and $E_{f}$ are the energy per unit of mass of the disk particle before and after the 
encounter, respectively. In a Keplerian potential, it holds that 
\begin{equation}
E=-\frac{G^{2}M_{1}^{2} (1-e^{2})}{2J^{2}},
\end{equation}
where $e$ is the orbital eccentricity. Before the encounter $e=0$ and thus $E_{0}\simeq -GM_{1}/(2a_{2})$.
After the encounter $e=A_{1}/R_{0}$, and therefore
\begin{equation}
E_{f}= -\frac{G^{2}M_{1}^{2}(1-A_{1}^{2}/R_{0}^{2})}{2(J_{0}+\delta J)^{2}}
\label{eq:Eff}
\end{equation}
Combining Eqs. (\ref{eq:amplitude})-(\ref{eq:Eff}), and noting that $\kappa_{0}=\Omega_{0}$ in
a Keplerian disk, we obtain
\begin{equation}
\delta J = -\frac{G^{2}M_{2}^{2}}{16\omega^{3}R_{0}^{4}}
\left(-2K_{0}\left[\frac{\xi_{0}}{2R_{0}}\right]+
\frac{\Delta_{0}}{\xi_{0}}K_{1}\left[\frac{\xi_{0}}{2R_{0}}\right]\right)^{2}.
\end{equation}

Since the time between successive encounters is $\simeq \pi/\omega$, the excitation torque
exerted on a differential ring of width $dR_{0}$, radius $R_{0}$ and surface density $\Sigma_{0}$, 
denoted by $dT_{g}/dR_{0}$, is $(\omega/\pi) 2\pi R_{0} \Sigma_{0} \, (\delta J)$.
Therefore,
\begin{equation}
\frac{dT_{g}}{dR}=-\frac{1}{8}q^{2}\Sigma \omega^{2} a_{2}^{3} \left(\frac{a_{2}}{R}\right)^{3}
\left(-2K_{0}\left[\frac{\xi}{2R}\right]+
\frac{\Delta }{\xi }K_{1}\left[\frac{\xi}{2R}\right]\right)^{2},
\label{eq:torqueKK_app}
\end{equation}
where we have dropped the subscripts on $R$, $\Delta$, $\Sigma$ and $\kappa$.

The present derivation of the exchange rate of angular momentum between
the secondary and the disk is valid only for encounters that produce
a small deflection of the orbit of the disk particle. A measurement of
the level of deflection of the orbit is the eccentricity $e$
induced in a disk particle by the passing perturber. The scattering
approximation should be valid provided that $e<0.1$. 
The condition $e=A_{1}/R_{0}<0.1$ implies that the present
approach is valid only for impact parameters $\Delta$ satisfying
\begin{equation}
\bigg|-K_{0}\left(\frac{\xi}{2R}\right)+\frac{\Delta}{2\xi} K_{1} 
\left(\frac{\xi}{2R}\right)\bigg| < 0.1 \left(\frac{\omega^{2} R^{3}}{GM_{2}} \right)= 0.1 q^{-1}.
\label{eq:condition1}
\end{equation}
In the particular case of a binary with $q=10^{-3}$,  the condition (\ref{eq:condition1}) is met for 
any value of $\Delta$ provided that $R_{\rm soft}\geq 0.005 a_{2}$.
On the other extreme, for $R_{\rm soft}=0$, the condition (\ref{eq:condition1}) can
be recast as
\begin{equation}
\frac{R_{\rm acc}}{\Delta}-\frac{q}{2}\ln\left(\frac{a_{2}}{\Delta}\right)
-0.405q\leq 0.05,
\end{equation}
where we have assumed $\xi\ll R\simeq a_{2}$.
For $q\leq q_{\rm cr}^{(r)}$, this implies that $\Delta\gtrsim 20R_{\rm acc}$.

\subsection{Spherical harmonic potential}
For comparison, instead of a Keplerian disk, now we consider a homogeneous background
mass distribution $\Phi_{b} (r)=\omega r^{2}/2$, with $\omega$ the constant angular
frequency. 
Following the same procedure, we find that the excitation torque density is
\begin{equation}
\frac{dT_{g}}{dR}=-\frac{1}{2}q^{2}\Sigma \omega^{2} a_{2}^{3} \left(\frac{a_{2}}{R}\right)^{3}
\left(-K_{0}\left[\frac{\xi}{R}\right]+
\frac{\Delta }{\xi }K_{1}\left[\frac{\xi}{R}\right]\right)^{2}.
\end{equation}
It is easy to show that the leading term in this case is the same as in the Keplerian disk.

\end{document}